\documentclass[12pt,a4paper,final]{iopart}
\usepackage{iopams}
\usepackage{amssymb}
\usepackage[square,numbers,sort]{natbib}  
\usepackage{graphicx}
\usepackage[breaklinks=true,colorlinks=true,linkcolor=blue,urlcolor=blue,citecolor=blue]{hyperref}
\usepackage{bm}

\expandafter\let\csname equation*\endcsname\relax

\expandafter\let\csname endequation*\endcsname\relax

\usepackage{amsmath}
\usepackage{url}

\begin{document}

\title[From quantum to thermal  topological-sector fluctuations of bosons in a ring lattice]{From quantum to thermal topological-sector fluctuations of strongly interacting bosons in a ring lattice}

\author{Tommaso Roscilde$^{1,2}$, Michael F. Faulkner$^{1,3}$, Steven T. Bramwell$^{3}$, and Peter C. W. Holdsworth$^{1}$ }
\address{$^1$ Laboratoire de Physique, CNRS UMR 5672, Ecole Normale Sup\'erieure de Lyon, Universit\'e de Lyon, 46 All\'ee d'Italie, 
Lyon, F-69364, France}
\address{$^2$ Institut Universitaire de France, 103 boulevard Saint-Michel, 75005 Paris, France}
\address{$^3$ London Centre for Nanotechnology and Department of Physics and Astronomy, University College London, 17-19 Gordon Street, London WC1H 0AJ, United Kingdom}
\ead{tommaso.roscilde@ens-lyon.fr}

\begin{abstract}
Inspired by recent experiments on Bose-Einstein condensates in ring traps, we investigate the topological properties of the phase of a one-dimensional Bose field in the presence of both thermal and quantum fluctuations -- the latter ones being tuned by the depth of an optical lattice applied along the ring. In the regime of large filling of the lattice, quantum Monte Carlo simulations give direct access to the full statistics of fluctuations of the Bose-field phase, and of its winding number $W$ along the ring. At zero temperature the winding-number (or topological-sector) fluctuations are driven by quantum phase slips localized around a Josephson link between two lattice wells,  and their { susceptibility} is found to jump at the superfluid-Mott insulator transition. At finite (but low) temperature, on the other hand, the winding number fluctuations are driven by thermal activation of nearly uniform phase twists, whose activation rate is governed by the superfluid fraction. A quantum-to-thermal crossover in winding number fluctuations is therefore exhibited by the system, and it is characterized by a conformational change in the topologically non-trivial configurations, from localized to uniform phase twists,  which can be experimentally observed in ultracold Bose gases via matter-wave interference.     
\end{abstract} 

\vspace{2pc}

\section{Introduction and main results}
 
 Ultracold atoms provide a very promising platform to investigate phenomena in mesoscopic physics, dominated by quantum coherence across the entire sample, and by strong fluctuations, both of quantum and thermal origin. Recent experiments have revealed the wide potential of \emph{atomic circuits}  to study quantum coherent transport \cite{Brantutetal2012,Stadleretal2012,Eckeletal2014-2,Krinneretal2015}, reproducing fundamental phenomena such as persistent currents, hysteresis, conductance quantization, etc. In particular ring traps \cite{Wrightetal2013,Eckeletal2014-2,Cormanetal2014,Eckeletal2014} are emerging as a fundamental platform mimicking electric coils, fundamentally sensitive to rotation or gauge fields, and they represent the simplest, topologically non-trivial geometry to investigate different topological sectors of a many-body model. 
 
A novel element introduced by cold-atom physics is the ability to study quantum transport phenomena with particles obeying bosonic statistics. In this context, a fundamental asset of cold-atom experiments is the use of matter-wave interferometry (or \emph{heterodyning}) to reconstruct the phase pattern of the many-body bosonic state -- an aspect which has been extensively used, among other things, to reveal the rapid spatial decay of first-order coherence in quasi-1d Bose gases \cite{Hofferberthetal2007,Hofferberthetal2008}, or topological defects in the phase pattern in quasi-2$d$ gases across a Kosterlitz-Thouless (KT) transition \cite{Hadzibabicetal2006}. Very recent experiments have probed the interference between a quasi-1d gas in a ring trap and a nearly uniform condensate \cite{Cormanetal2014, Eckeletal2014}, within a geometry which allows the full reconstruction of the phase pattern along the ring \cite{Mathewetal2015}. In particular the ring geometry represents the simplest, non-simply connected platform to investigate topological defects in the phase pattern, in the form of quantized vortices 
\begin{equation}
\oint {\bm \nabla} \phi \cdot d{\bm l} = 2\pi W ~~~~~~~~~~ W=0, \pm 1, \pm 2, ... 
\label{e.vortex}
\end{equation}
where $\phi({\bm r})$ is the phase of the Bose field, and $W$ is the winding number of the phase pattern. 
The winding number partitions the Hilbert space of bosonic fields on a ring into different topological sectors. Recent experiments have shown the ability to generate finite winding numbers via a thermal quench  \cite{Cormanetal2014} or by stirring the condensate via rotation of a potential barrier along the ring \cite{Wrightetal2013,Beattieetal2013,Eckeletal2014-2,Eckeletal2014}; as well as the ability to faithfully reconstruct the whole phase profile along the ring and its associated winding number by matter-wave heterodyning \cite{Cormanetal2014,Eckeletal2014}. 

 We draw inspiration from these experiments to investigate phase coherence and winding number fluctuations in a regime  unexplored so far in the experiments, namely that of strong correlations among the bosons. Strong correlations are most naturally induced in cold-atom setups using optical lattices \cite{Blochetal2008}, which can destroy condensation by driving the system towards a Mott insulating state. An optical lattice can be created experimentally along the ring trap by either using Laguerre-Gauss beams with high angular momentum \cite{Amicoetal2005} or via spatial light modulators \cite{Cormanetal2014, Corman_private_communication, FrankeArnoldetal2007, Aghamalyanetal2015}. In this way the ring is partitioned into potential wells separated by tunnel barriers, forming a chain of bosonic Josephson junctions; and the tunnel barriers can act as nucleation centers for phase twists in the Bose field. 
 
  In the presence of finite interactions, phase twists can be generated via quantum fluctuations of the local phase of the Bose field, due to the duality between phase and density, $[\phi,n]=i$. Given the vortex-quantization condition in Eq.~(\ref{e.vortex}), interactions can generate quantum coherent fluctuations between different winding numbers $W$ by inducing $\pm 2\pi$ revolutions in the phase along the ring, called \emph{quantum phase slips} (QPS). QPS can be regarded as the dual phenomenon to Josephson tunneling \cite{LikharevZ1985}, as they establish coherence between Bose-field configurations with a well defined phase, dual to the coherence between Fock states established by Josephson tunneling.  Due to their fundamental importance, QPS have been the subject of intense research in superconducting nanostructures, including narrow superconducting wires \cite{Astafievetal2012} and Josephson junction arrays \cite{Popetal2010,Manucharyanetal2012,Erguletal2013}, as well as cold atoms trapped in arrays of optical-lattice tubes \cite{Tanzietal2013} and Helium trapped in nanostructures \cite{Hoskinsonetal2006}. Phase slips (either of thermal or quantum origin) are responsible for the decay of supercurrents in quasi-one-dimensional superconducting devices and confined superfluids, and their detection so far is primarily based on transport measurements. On the other hand, the realization of QPS in ring-trapped condensates clearly opens the possibility of detecting them interferometrically using the same scheme as in the recent experiments reported in Refs.~\cite{Cormanetal2014, Eckeletal2014}. But how can one interferometrically distinguish a QPS from a phase slip of thermal origin (or thermal phase slip, TPS)? And under which conditions does one cross over from a regime dominated by thermal winding number fluctuations, to a regime dominated by quantum winding number fluctuations induced by QPS?
  
   These are precisely the questions addressed by our present paper. Based on an extensive numerical study of the quantum phase model -- faithfully representing the physics of optical lattices in the large filling regime -- we reconstruct the quantum-to-thermal crossover in winding number fluctuations across the phase diagram of the system, namely from the superfluid phase at weak coupling to the Mott insulating phase at strong coupling. In particular our findings show that QPS are localized objects, occurring in an uncorrelated fashion in different segments of the ring, and the ensuing quantum winding number fluctuations obey Poissonian statistics, growing linearly with system size. The measurable interferograms manifest the localized nature of QPS in the form of line dislocations in the interference fringes, akin to those caused by a tunnel barrier \cite{Eckeletal2014}. Our data strongly suggest that the { susceptibility} of quantum winding-number fluctuations exhibits a jump at the quantum phase transition from superfluid to Mott insulator, endowing such a transition with a precise topological nature. 
   
   On the other hand, at finite (but sufficiently low) temperatures thermal winding number fluctuations  occur between the lowest energy states in each topological sector, implying that the $2\pi W$ phase twist characterizing such sectors is rather uniformly distributed along the ring. A uniform phase twist translates into measurable interferograms whose fringes form regular spirals, similar to those observed after thermal quenches \cite{Cormanetal2014}. The crossover from localized QPS to extended TPS turns out to be very sharp for sufficiently weak interactions among the bosons, whereas it is broad (or even poorly defined) in the strongly interacting regime.  The thermal activation of winding-number fluctuations is governed by the superfluid density at zero temperature, and hence it may serve as a way to measure this quantity, so far elusive in cold-atom experiments.
    
  Our paper is organized as follows: Sec.~\ref{s.QPM} describes the quantum phase model and the path-integral approach used in the numerical calculations, the path-integral picture of thermal and quantum phase slips, and the basics of their experimental detection via matter-wave interferometry; Sec.~\ref{s.quantum} focuses on the thermal regime of winding-number fluctuations and on the superfluid/Mott-insulator transition;  Sec.~\ref{s.thermal} discusses the thermal onset of winding-number fluctuations and its relationship to the superfluid fraction; the crossover between the quantum and the thermal regime, as well as a peculiar thermal revival of quantum fluctuations above the superfluid ground state, is the subject Sec.~\ref{s.crossover}; finally conclusions are drawns in Sec.~\ref{s.conclusions}.

\section{Quantum phase model and its topological excitations}
\label{s.QPM}

\subsection{From Bose-Hubbard to quantum phase model}
\label{s.XYmapping}

 A faithful description of ultracold bosons in an optical lattice is provided by the Bose-Hubbard model 
 \begin{equation}
 {\cal H} = -J\sum_{i=1}^L \left (b_i^{\dagger} b_{i+1} + {\rm h.c.} \right) + \frac{U}{2} \sum_i n_i(n_i-1)
 \end{equation}
 where $b_i$, $b_i^{\dagger}$ are Bose field operators defined on a ring of length $L$ with periodic boundary conditions, $b^{(\dagger)}_{L+1}=b^{(\dagger)}_1$, { $J$ is the hopping energy and $U$ the on-site repulsion}. 
 Hereafter we assume an average \emph{integer} filling $\bar{n} = \langle n_i \rangle$ of the chain.  
 Strictly working with Bose fields turns out to be impractical when one aims at investigating the quantum or thermal fluctuations of the phase pattern in the Bose field itself: indeed the phase operator is not well defined, unless one works in the large-filling regime $\bar{n} \gg 1$. In this limit one may map the Bose operators to amplitude and phase operators and neglect amplitude fluctuations, namely one can take $b_i \approx \sqrt{\bar{n}} ~e^{i\phi_i}$ and $n_i \approx -i \frac{\partial}{\partial \phi_i}$. This mapping reduces the Bose-Hubbard Hamiltonian to a quantum phase (or quantum rotor) model  (QPM) -- a cornerstone in the study of Josephson junction arrays \cite{Fazioetal2001}:
 \begin{equation}
 {\cal H} = -2J\bar{n} \sum_i \cos(\phi_i-\phi_{i+1}) - \frac{U}{2} \sum_i \frac{\partial^2}{\partial\phi_i^2} ~.
 \end{equation}
 { In the following we shall express all energy scales in units of $2J{\bar n}$, namely we introduce the reduced interaction $u = U/(2J{\bar n})$ and reduced temperature $t = k_B T/(2J{\bar n})$.}
 The QPM model is perfectly suited to the study of the quantum dynamics of the Bose-field phase in a regime in which density fluctuations are irrelevant -- which is at the center of our investigation. 
 As we shall see later (Sec.~\ref{s.visual}) the QPM regime of large filling is also most appropriate in the perspective of extending the existing experimental setups and detection schemes for ring-trapped condensates \cite{Cormanetal2014,Eckeletal2014} to the regime of strong correlations.   
  
 The path-integral representation of the partition function of the QPM maps the latter model onto an effective classical XY model in (1+1) dimensions \cite{Wallinetal1994,Sondhietal1997}, with a spatially anisotropic Hamiltonian
 \begin{equation}
 { {\cal H}_{\rm eff} =\sum_{i=1}^{L} \sum_{k=1}^{M} \left [ -\frac{2J \bar n}{M}  \cos(\phi_{i,k}-\phi_{i+1,k}) - \frac{2 M}{\beta^2 U} \cos(\phi_{i,k}-\phi_{i,k+1}) \right ]}
 \label{e.XY}
   \end{equation}
   defined on a $L\times M$ torus, where $M$ is the size of the grid in the imaginary-time dimension $\tau\in[0,\beta]$ ($\beta=1/(k_B T)$), discretized into infinitesimal steps $\delta \tau = \beta/M$.

 The exact properties of the QPM are strictly recovered in the $M\to\infty$ limit, due to the Trotter-Lie decomposition at the heart of the path-integral approach \cite{Sondhietal1997}. Moreover the mapping of the interaction term to an XY interaction in imaginary time is accurate provided that the dimensionless parameter $\epsilon = \beta U/M$ is small \cite{Wallinetal1994}. Hence $\epsilon\ll 1$ is the most stringent requirement for the quality of the quantum-to-classical mapping. In the following we use $\epsilon = 10^{-2}$, which essentially guarantees convergence of all the quantities of interest towards their $M\to\infty$ limit. 
 
{ With $\epsilon$ held fixed, the effective XY Hamiltonian (divided by the temperature scale) is most conveniently recast in the form
\begin{equation}
 \beta{\cal H}_{\rm eff} = \sum_{i=1}^{L} \sum_{k=1}^{M} \left [ -K \cos(\phi_{i,k}-\phi_{i+1,k}) - K_{\tau}  \cos(\phi_{i,k}-\phi_{i,k+1}) \right]
 \label{e.XY2}
 \end{equation}
where we have introduced the coupling constants $K = \epsilon/u$ and $K_{\tau} = 2/\epsilon$. It is important to remark that the coupling constants are independent of temperature, and they are also held fixed for a given fixed strength of the interaction; hence varying the temperature has the unique effect of controlling the length of the extra dimension, $M = u/(t\epsilon)$. On the other hand, changing the interaction strength $u$ at fixed temperature has the effect of changing both the length $M$, and the anisotropy between the space-like ($K$) and imaginary-time-like ($K_{\tau}$) couplings.}

 The quantum-to-classical mapping of the QPM opens the path to studying the equilibrium properties of the system via Monte Carlo simulations. We perform simulations on chains of length between $L=24$ and $L=216$, and for a broad range of values for the ratio $U/J$, covering the superfluid/Mott-insulator (SF/MI) transition. We make use of a combined update scheme comprising single-spin Metropolis updates; overrelaxation moves; Metropolis updates proposing rotations of entire chains of spins in imaginary time (which essentially behave as gigantic spins in the limit $U\to 0$); and Wolff cluster updates \cite{Wolff1989}.  In the following we shall denote with $\langle ... \rangle_{\rm MC}$ the Monte Carlo averages, calculated from the effective classical XY model of Eq.~(\ref{e.XY}), to contrast it with the general statistical averages of quantum operators, denoted with $\langle ... \rangle$. 
 
 The path-integral Monte Carlo approach to the QPM is particularly appealing, as the computational basis diagonalizes the phase operator, and hence one has readily access to the statistics of fluctuations -- both thermal and quantum (coherent)-- of the phase pattern, and in particular to its winding number. 
 Indeed each imaginary-time slice in the path-integral representation samples a possible phase pattern, which in turn is projectively measured in an interferometric experiment.  
 
The winding number of the phase pattern at the $k$-th imaginary-time slice is defined as
\begin{equation} 
W_k = \sum_i F(\phi_{i+1,k}-\phi_{i,k})
\end{equation}
where
\begin{equation}
F(\Delta\phi)= 
\begin{cases} \Delta\phi  & ~~~~~ \text{if}~~  |\Delta\phi|\le \pi  \\ 
 2\pi-\Delta\phi   & ~~~~~ \text{if}~~ \Delta\phi \geq \pi \\ 
  2\pi+\Delta\phi & ~~~~~ \text{if}~~ \Delta\phi \leq -\pi ~.
   \end{cases}   
   \label{e.F}
 \end{equation}
 
  \begin{figure}[htb!]
 \centering
  \mbox{
  \includegraphics[width=5cm]{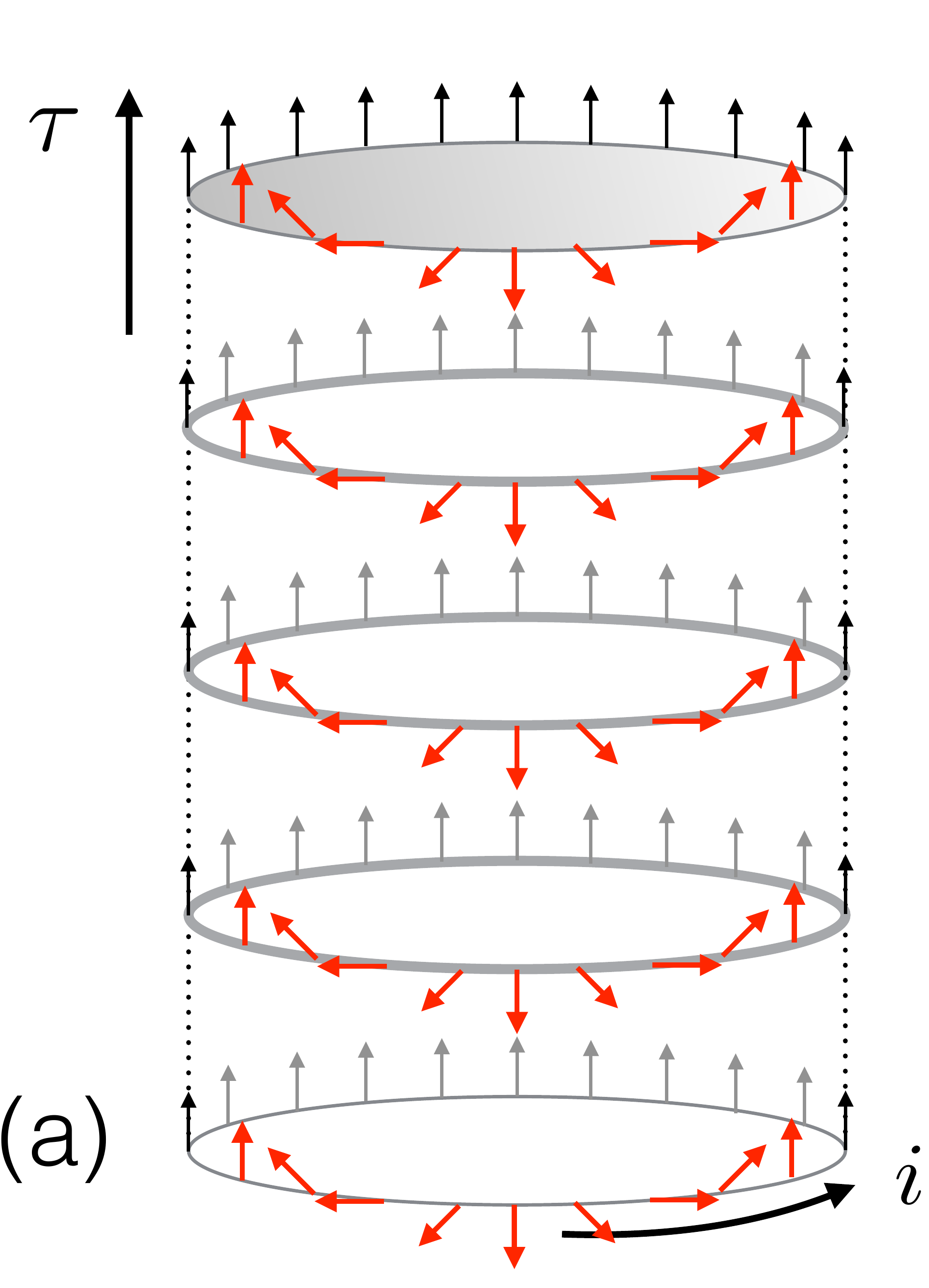}~~~~
  \includegraphics[width=5cm]{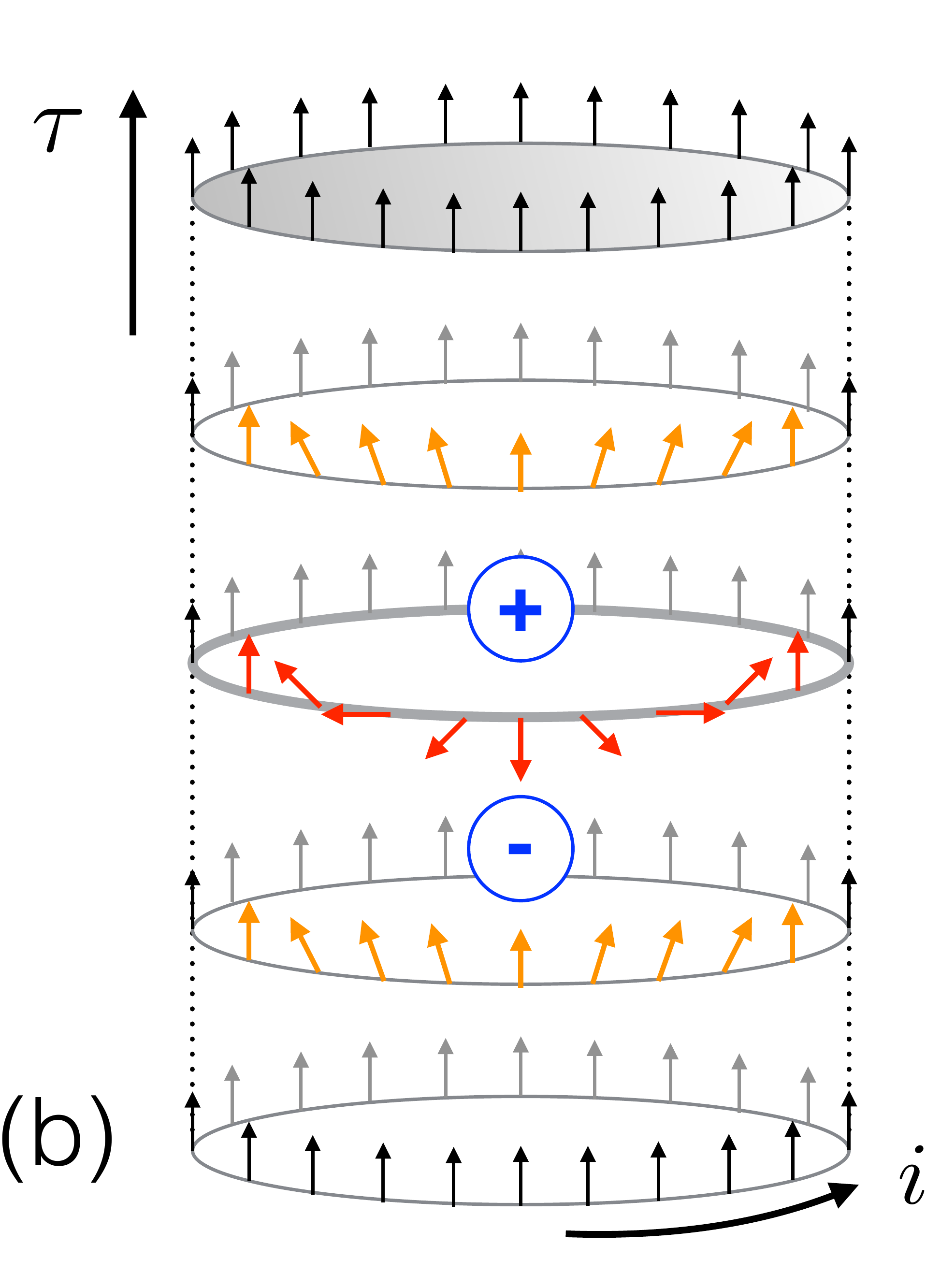}~~~~
  \includegraphics[width=5cm]{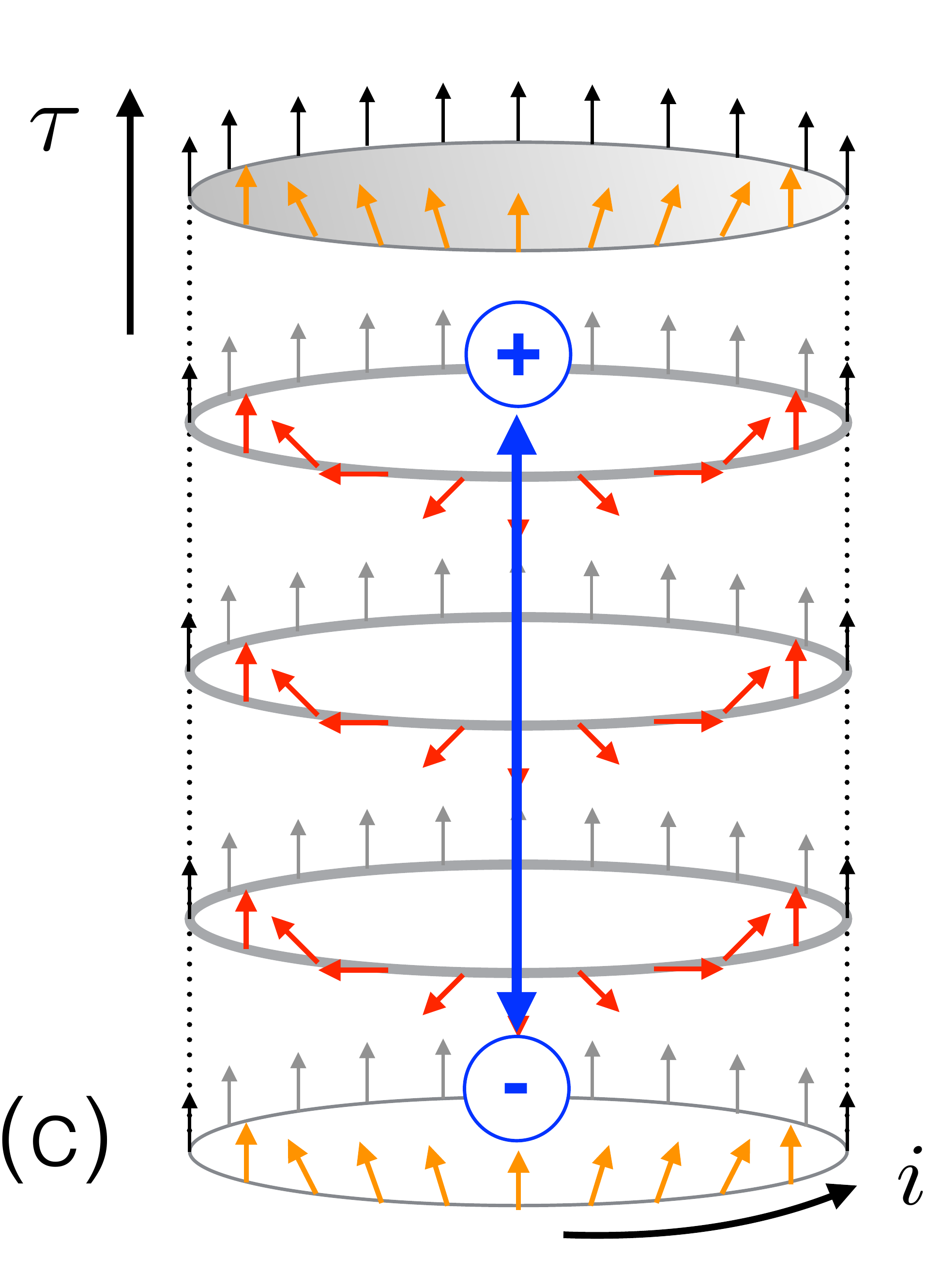}}
 \caption{(a) Thermal phase slip (TPS) induced by incoherent thermal fluctuations; (b) Quantum phase slip (QPS), localized in the imaginary time direction; (c) QPS after unbinding of vortex-antivortex excitations along the imaginary-time direction.}
\label{f.PI}
\end{figure}

\subsection{Thermal vs. quantum phase slips in the path-integral representation} 
\label{s.PI}
 
 As already mentioned in the introduction, bosonic chains can support two different types of topological excitations, leading to finite winding-number configurations, namely quantum and thermal phase slips (QPS and TPS). The two different kinds of excitations are best visualized within the path-integral formalism, as sketched in Fig.~\ref{f.PI}. 
 
  TPS are common to a classical XY chain, namely they correspond to the appearance of a finite winding number $W$ on all imaginary-time slices -- see Fig.~\ref{f.PI}(a). This requires a massive rearrangement of the path-integral configuration, which can be achieved in a Monte Carlo simulation via Wolff clusters percolating along the imaginary-time direction, producing incoherent winding number fluctuations on all imaginary-time slices.  
 
 The path-integral formalism offers an ideal tool to visualize QPS. Indeed the mapping of the $d=1$ QPM onto a $(1+1)$-dimensional XY model as in Eq.~(\ref{e.XY}) allows one to identify QPSs in the space-time phase pattern $\{\phi_{i,k}\}$ as those configurations in which the imaginary-time evolution leads to a change in winding number between different imaginary-time slices $k$, see Fig.~\ref{f.PI}(b). Such a situation is only possible if two space-time topological excitations, in the form of a vortex-antivortex (V-AV) pair, have appeared in the system. 
  In the presence of a finite interaction among particles, QPS originating from V-AV pairs displaying a finite separation along the imaginary-time axis will always exist in the ground state. { Indeed, unless $u=0$, the imaginary-time direction has a finite (namely, non-zero) length $M$ (at fixed $\epsilon$) and a finite (namely, not infinite) coupling constant $K_{\tau}$ along that direction: hence the effective XY model in Eq.~(\ref{e.XY}) always displays nucleation of V-AV pairs separating in imaginary time -- and this property persists down to zero temperature.} Below the Kosterlitz-Thouless SF-MI transition, the V-AV pairs are bound, implying that the QPS are localized both in real space and imaginary time, namely they extend over an imaginary-time and spatial width which are not scaling with { the corresponding sizes $M$ and $L$}, respectively. In this regime, the quantum winding number fluctuations are of ``virtual" nature, namely the $W=0$ topological sector is only weakly admixed with the $W\neq 0$ sectors via localized QPS. In the path-integral image, this correspond to short-lived (in imaginary time) fluctuations towards $W\neq 0$ values. Entering into the MI phase, the unbinding of V-AV pairs produces extended QPS along the imaginary-time direction, which induce strong coherent quantum fluctuations of the winding number. The separation between vortex cores may now scale { with the length of the imaginary-time direction and with system size}; in particular the vortex cores may wind around the imaginary-time direction, inducing a global change in the winding number \cite{Faulkneretal2015} akin to the thermal activation of a TPS, but of coherent nature.

 \begin{figure}[htb!]
 \centering
  \includegraphics[width=12cm]{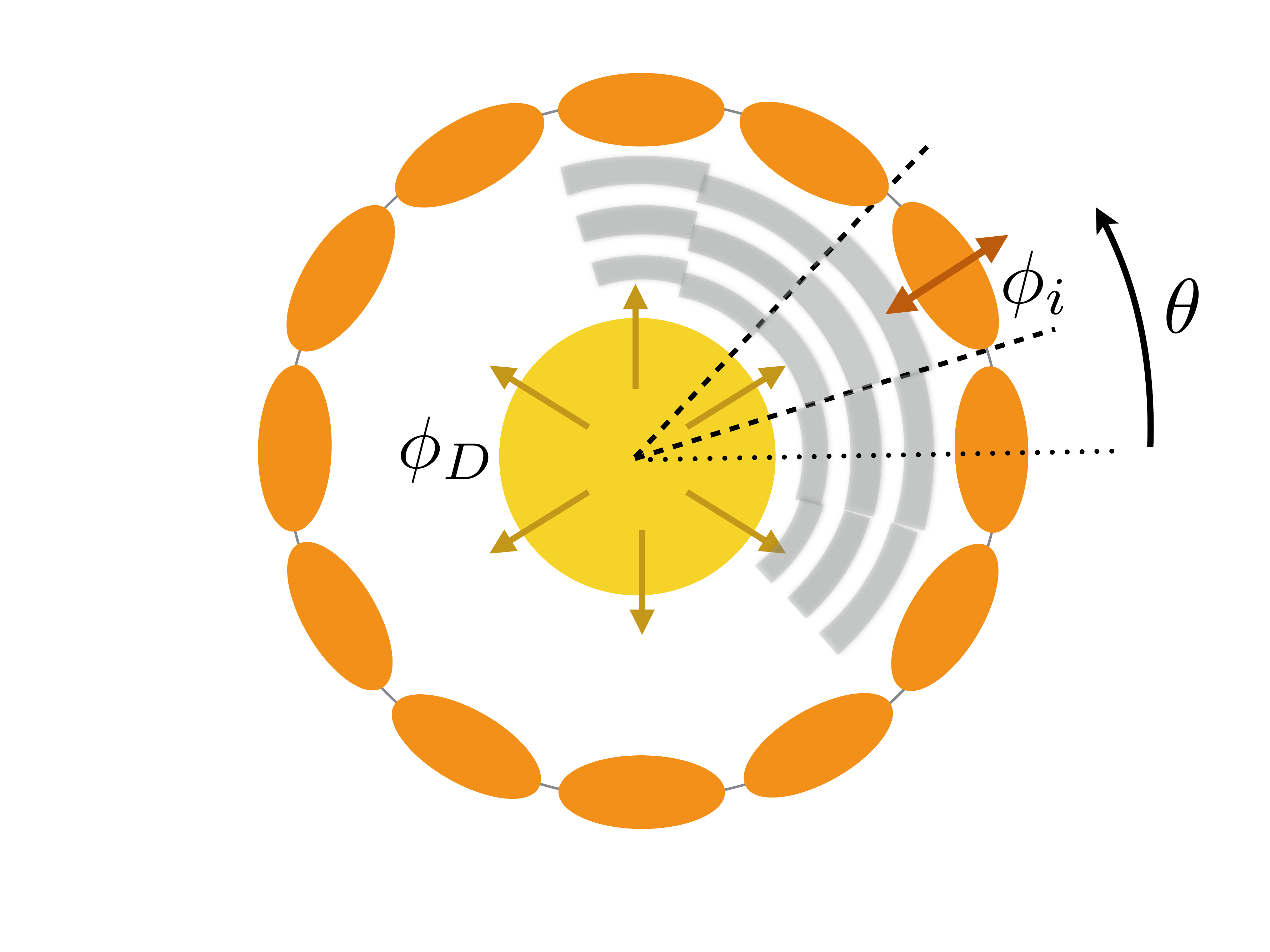}~~~~
 \caption{Interferometric scheme to reconstruct the phase pattern of the Bose field along the ring lattice. After release of the trapping potentials in the plane of the drawing, the rapid expansion along the radial direction gives rise to interference among the disk condensate (yellow disk) and the minicondensates along the ring (orange ellipses), with the appearance of fringes (grey/white stripes) whose spatial phase reflects the phase difference $\phi_i - \phi_D$.}
\label{f.interf}
\end{figure}

\subsection{Visualizing topological defects via matter-wave heterodyning} 
\label{s.visual}

 Recent experiments in ring-trapped Bose-Einstein condensates \cite{Cormanetal2014,Eckeletal2014} have shown the unique ability to reconstruct the phase pattern along the ring, making use of interference with a condensate trapped at the center of the ring in a disk-shaped potential. In the following we shall imagine that such a scheme is equally applicable to the case in which an optical lattice has been created along the ring, dividing it into a sequence of (single-mode) mini-condensates with filling $\bar{n} \gg 1$ -- obtained under the condition of the optical-lattice period being significantly larger than the interparticle spacing. Moreover, as suggested by our sketch in Fig.~\ref{f.interf}, we shall imagine that the optical lattice wells are oblate along the tangential direction to the ring, so that the strongest confinement remains along the radial direction.      
 
 Under the above conditions, releasing the potentials in the plane of the ring produces a quasi-2$d$ expansion of each mini-condensate, as well as of the disk-trapped condensate, which is much faster along the radial direction than along the polar one. As is well known, the interference pattern among ``two condensates that have never seen each other" \cite{Anderson} is akin to that of two gauge-symmetry-breaking matter waves with a well defined phase but a random phase relationship -- which can be ascribed, as a convention, to an arbitrary phase $\phi_D$ of the condensate in the disk. For sufficiently short time scales the expansion can be approximated as being purely radial, and the resulting density profile along the radial direction at a given polar angle $\theta$ can be obtained approximately from the far-field radial expansion of both the matter wave in the ring and the disk. After a time of flight $t_{\rm TOF}$ the ring wavefunction and the disk wavefunction in the wedge centered around the angle $\theta$ (see Fig.~\ref{f.interf}) read
\begin{eqnarray}
 \psi_R(r,\theta) & \sim &  \tilde\psi_{\perp,R}(k_r) ~e^{i\frac{m(r-d)^2}{2\hbar t_{\rm TOF}} } ~e^{i\phi_{i(\theta)}} \nonumber \\
 \psi_D(r,\theta) & \sim &  \tilde\psi_{\perp,D}(k_r) ~e^{i\frac{m r^2}{2\hbar t_{\rm TOF}} } ~e^{i\phi_D}
\end{eqnarray} 
 where $\tilde\psi_{\perp,R}$ and  $\tilde\psi_{\perp,D}$ are the (real-valued) Fourier transform of the transverse profiles of the ground-state wavefunction in the potential well and in the disk respectively, calculated in $k_r =\frac{mr}{\hbar t}$; $\phi_{i(\theta)}$ is the phase of the mini-condensate in the $i$-th lattice well, corresponding to the polar angle $\theta$; $d$ is the ring radius; and $m$ is the mass of the particles. The interference term then takes the form
 \begin{equation}
 2{\rm Re} [\psi_R(r,\theta) \psi_D^*(r,\theta)] = 2 ~\tilde\psi_{\perp,R}(k_r) \tilde\psi_{\perp,D}(k_r) \cos\left[ Q r - (\phi_{i(\theta)}-\phi'_D) \right]
 \label{e.interf}
 \end{equation}
 where $Q = \frac{md}{2\hbar t_{\rm TOF}}$ and $\phi'_D = \phi_D + \frac{md^2}{2\hbar t_{\rm TOF}}$.
 
 Putting together the different interference patterns developing along the radial direction results in a global pattern equivalent to the one partially sketched in Fig.~\ref{f.interf}, in which the fringe spacing (controlled by the $Q$ vector) is the same at all polar angles $\theta$, but the fringe pattern along the radial direction is shifted by the phase profile $\phi_{i(\theta)}$ along the ring. This direct correspondence between the Bose-field phase and the spatial phase of the fringe pattern allows the full reconstruction of the phase structure, as done in Refs.~\cite{Cormanetal2014,Eckeletal2014}.  
 
  In the following sections we shall plot the interference term of Eq.~\eqref{e.interf} -- without the modulation due to the Fourier transforms of the disk and ring transverse modes  $\tilde\psi_{\perp,R(D)}$ -- to illustrate the contrasting experimental signatures of TPS and QPS.   
 
 \begin{figure}[htb!]
 \includegraphics[width=\columnwidth]{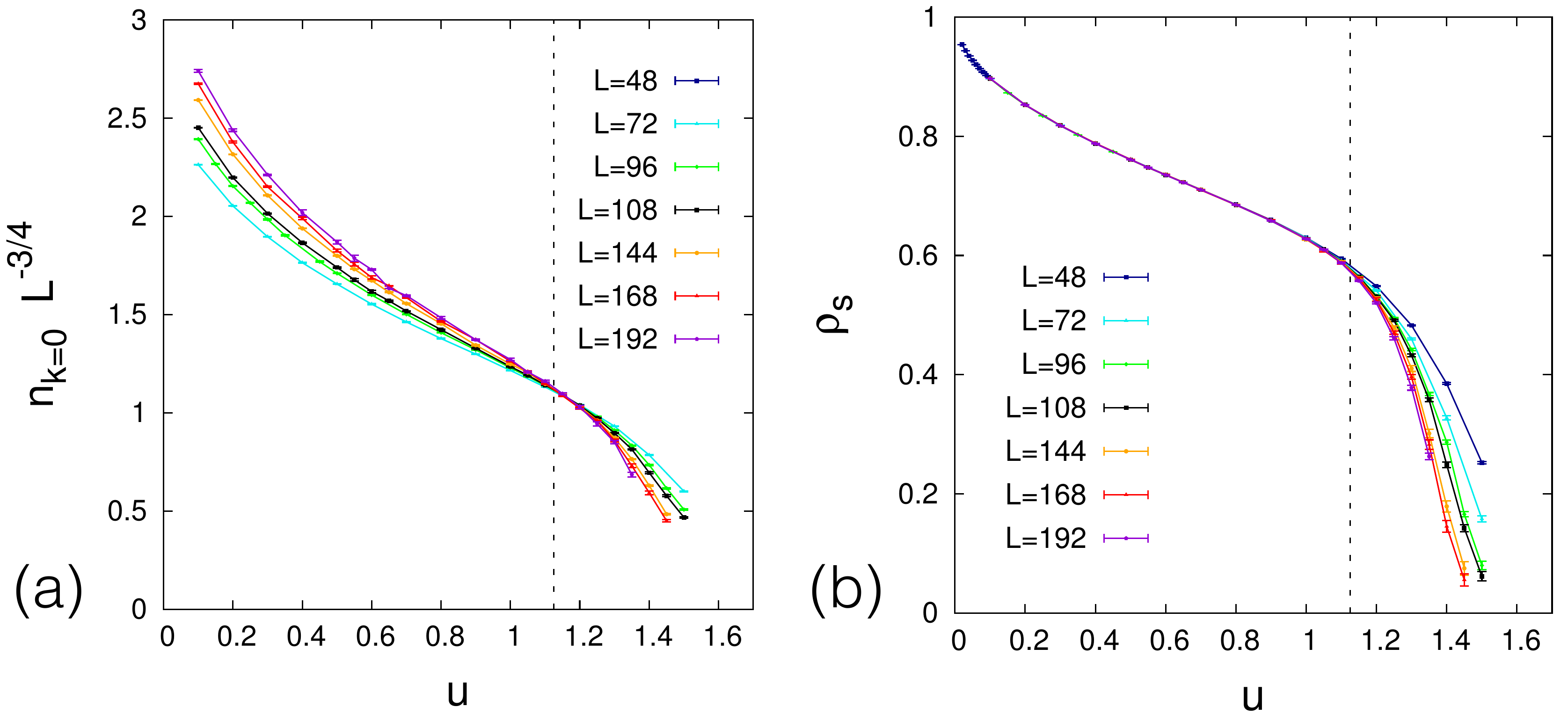}
 \caption{(a) Scaling plot of the momentum-distribution peak $n_{p=0}$ at temperature $t=1/L$; the crossing point marks the SF-MI quantum phase transition; (b) superfluid density for various system sizes. }
\label{f.SFMI}
\end{figure}

\section{Quantum regime: quantum phase slips}
\label{s.quantum}

\subsection{Superfluid-insulator transition}

 We begin the description of our results the from ground-state behavior at $T=0$. Finite-temperature effects are essentially removed from the Monte-Carlo simulations below a size-dependent temperature $t = 1/L$, which guarantees that the gapless phonon-like excitations in the superfluid phase are not thermally excited.   
 Fig.~\ref{f.SFMI} shows the determination of the SF/MI transition at $T=0$ from the Monte Carlo calculation of the $p=0$ population in the momentum distribution
 \begin{equation}
 \frac{n_{p=0}}{\bar{n}} = \frac{1}{ML} \sum_{ij} \sum_k \left \langle e^{i(\phi_{i,k} - \phi_{j,k})} \right \rangle_{\rm MC}~  
 \end{equation}  
 which is expected to scale as $L^{3/4}$ at the KT quantum critical point separating the SF from the MI. The crossing point of the different curves for $n_{p=0}~L^{-3/4}$ gives an estimate of $u=u_c\approx 1.12(5)$ for the critical point of the SF/MI transition.  
 This estimate is in very good agreement with the prediction of \cite{DanshitaP2011}, giving $u_c \approx 1.08$ for the Bose-Hubbard model in the limit of very large, integer filling. 
 At the same time the transition is marked by the vanishing of the superfluid fraction
 \begin{equation}
 \rho_s = \frac{1}{2J\bar{n}L} \frac{\partial ^2 F}{\partial \varphi^2} \Big |_{\varphi=0} = e - \frac{1}{L} \chi_J
 \label{e.rhos}
 \end{equation}
  where $F$ is the free energy of the system, and $\varphi$ is a phase twist applied in the space direction. Moreover we have denoted by
 \begin{equation}
 e = \langle \cos(\phi_i - \phi_{i+1})\rangle = \frac{1}{LM} \sum_{i,k} \left\langle \cos(\phi_{i,k} - \phi_{i+1,k}) \right\rangle_{\rm MC}
 \end{equation}
the average Josephson coupling term, and by 
 \begin{equation}
   \chi_{J} = \int d\tau~ \langle J(\tau) J(0) \rangle =    \frac{1}{t} \left \langle \left [ \frac{1}{M}\sum_{i,k} \sin(\phi_{i,k}-\phi_{i+1,k}) \right ]^2 \right \rangle_{\rm MC}   
 \label{e.TA}
 \end{equation}
the current susceptibility. The identities between the general definition and the Monte Carlo estimator of the above quantities are strictly recovered in the $M\to\infty$ limit. 
 Fig.~\ref{f.SFMI} shows that the SF/MI transition is marked by a strong suppression of the superfluid density calculated on a finite size, alluding to the presence of a jump expected for this quantity in the thermodynamic limit. 
 
  \begin{figure}[htb!]
 \centering
\includegraphics[width=\columnwidth]{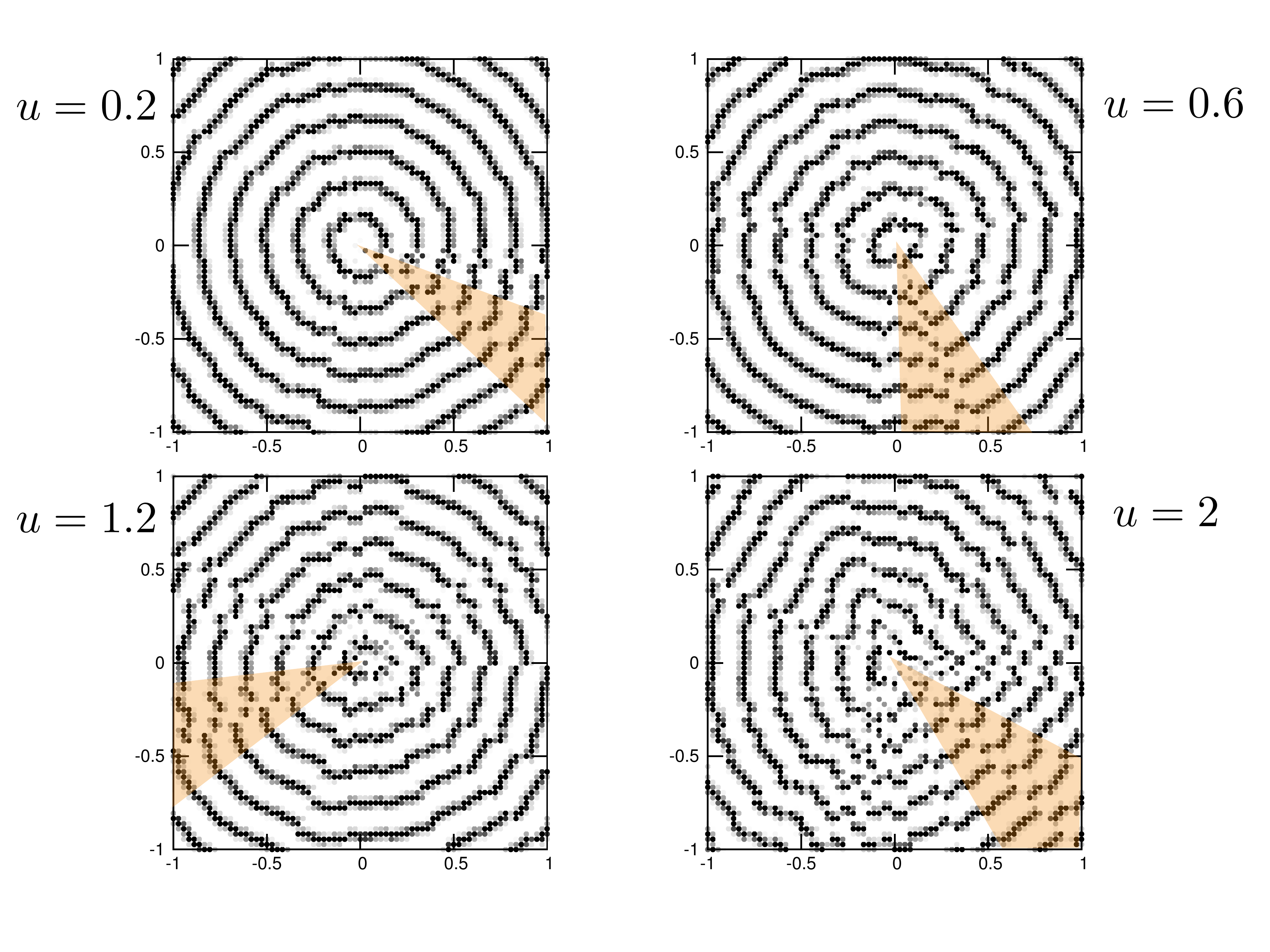}
 \caption{Inteference patterns with $|W|=1$, resulting from snapshots of the QMC simulation on a ring with L=72 at physical zero temperature and varying interaction strength. The plot represents the function $f(r,\theta) = 1+\cos(Qr+\phi_{i(\theta)})$ (see text), with logarithmic gray scale to enhance the fringe geometry; the darker regions correspond to higher density. The shaded areas mark the position of the localized QPS. }
\label{f.QPS}
\end{figure}

 \begin{figure}[htb!]
 \centering
\includegraphics[width=\columnwidth]{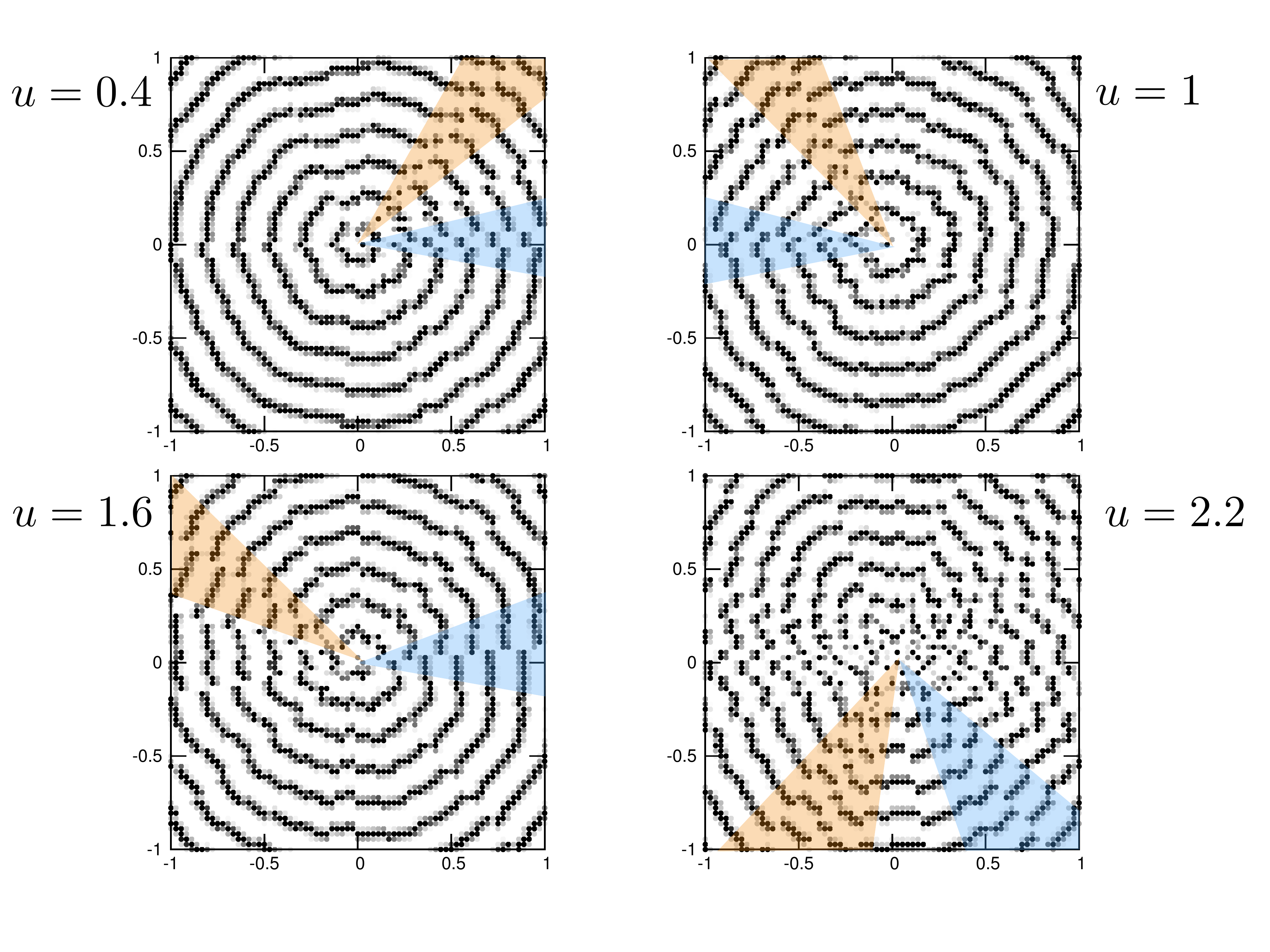}
 \caption{Inteference patterns with $|W|=2$, resulting from snapshots of the QMC simulation on a ring with L=72 at physical zero temperature and varying interaction strength. The plotted function and color scale are the same as in Fig.~\ref{f.QPS}.  The shaded areas mark the positions of the two localized QPS.}
\label{f.QPS2}
\end{figure}

\subsection{Winding-number fluctuations and quantum phase slips}

The winding-number fluctuations at $T=0$ are uniquely induced by QPS, which, as seen in the path-integral picture in Sec.~\ref{s.QPM}, are related to the nucleation of V-AV pairs separating in the imaginary-time dimension.
Figs.~\ref{f.QPS} and \ref{f.QPS2} show the reconstructed interference pattern from representative configurations $\{\phi_{i,k}\}$ with $i=1,...,L$ at fixed $k$, produced by the QMC simulation, and possessing a finite winding number $W$ ($|W|=1$ and 2).
In the SF regime of strong V-AV binding, the cores of the V-AV pairs remain very close to each other, thereby producing localized QPS in the phase configurations with $W\neq 0$; such localized QPS produce very sharp fringe dislocations in the interference patterns for weak interactions $u$, gradually spreading over an increasingly large range when $u$ increases to cross the SF/MI transition. The localized fringe dislocations reorganize the fringe pattern into a spiral, similarly to what is observed in the case of a ring with a tunnel barrier in Refs.~\cite{Eckeletal2014,Mathewetal2015} acting as a phase slip center; in our case of a Josephson junction chain, tunnel barriers separate each pair of sites, and QPS spontaneously nucleate at any position in the lattice.

   \begin{figure}[htb!]
 \centering
 \includegraphics[width=\columnwidth]{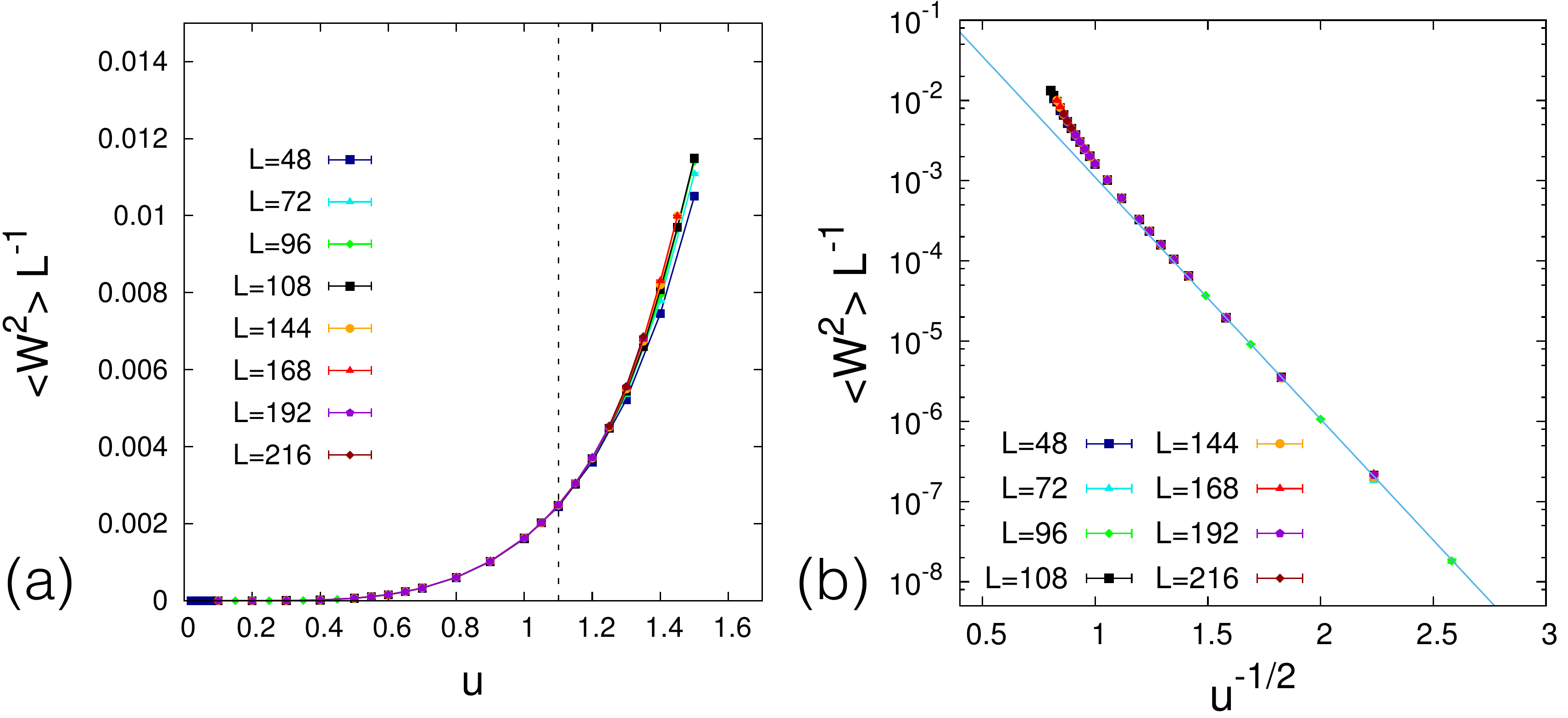}
 \caption{Quantum winding-number fluctuations at $T=0$. (a) Scaling of the variance of the fluctuations as a function of the interaction strength; the vertical dashed line marks the SF/MI transition; (b) Same plot as in (a), but as a function of $u^{-1/2}$; the solid line is a fit to the for $a\exp(-b/\sqrt{u})$ with $a \approx 1.15$ and $b\approx 6.95$.}
\label{f.W2}
\end{figure}

    \begin{figure}[htb!]
 \centering
 \includegraphics[width=0.8\columnwidth]{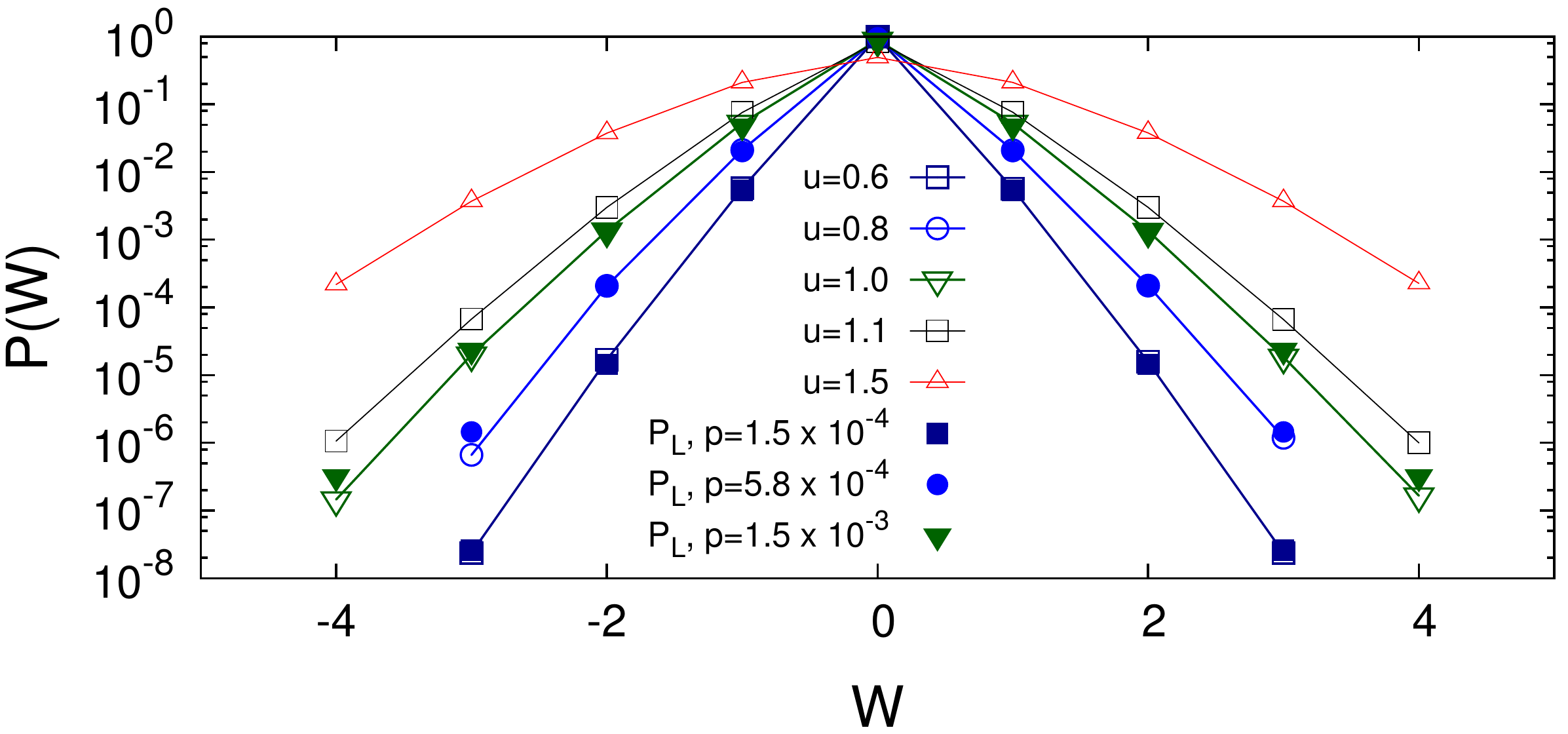}
 \caption{Probability distributions $P(W)$ of winding-number fluctuations in the ground state of a Bose ring lattice with $L=72$ and variable interaction parameter $u$ (only values $|W|\leq 4$ have been samples). The full symbols represent fits to the Ansatz in Eq.~\eqref{e.PLW}, only dependent on the probability $p$.}
\label{f.PW}
\end{figure}

\subsubsection{Superfluid phase}
 The localized nature of the QPS, and their diluteness in the confined regime $u\leq u_c$, makes them statistically independent: every tunnel barrier has a finite probability $p$ of nucleating a QPS with an arbitrary sign $\pm 2\pi$, and QPS sum up independently to produce a total winding number $W$ which, to a first approximation, is a Poissonian process with zero average and fluctuations 
 \begin{equation}
 \langle W^2 \rangle \approx p L~.
 \label{e.quantumregime}
 \end{equation}
 Such a linear scaling is indeed verified in the SF regime, as shown in Fig.~\ref{f.W2}(a).  
 A more refined analysis implies the necessity to consider that each configuration may display a number of QPS which follows a Poissonian distribution with parameter $\lambda=pL$;  { if $q$ is the number of QPS, each of them may display $\pm 2\pi$ phase slip at random}, resulting in a $W$ which is the displacement of a symmetric random walk with $q$ steps. The resulting probability is then that of the displacement of $W$ steps of a random walk of stochastic length $q$, namely
 \begin{equation}
 P_L(W) \approx e^{-pL}   \sum_{q=0}^{\infty} \frac{1}{\left(\frac{q+W}{2}\right)! \left(\frac{q-W}{2}\right)!} \left(\frac{pL}{2} \right)^q 
 \label{e.PLW}
 \end{equation} 
 This Ansatz is indeed well verified by our QMC data, as shown in Fig.~\ref{f.PW}. Fits to Eq.~\eqref{e.PLW} provide estimates for the probability $p(u)$ which are in very good agreement with those drawn from the scaling plot in Fig.~\ref{f.W2}(a).

 
 The probability $p$ to nucleate a QPS on a given tunnel barrier can be estimated by analyzing the Hamiltonian of a single Josephson junction (JJ), namely 
 \begin{equation}
 {\cal H}_{JJ} = - E_J \cos\phi - 4E_c \frac{d^2}{d\phi^2}
 \label{e.JJ}
 \end{equation}
  where $\phi$ is the phase difference across the junction,  $E_J = 2J\bar{n}$ is the Josephson energy, and $E_c = U/8$ the charging energy. Following Ref.~\cite{LikharevZ1985} one can approximate $\phi$ as the (unbounded) position of a particle, so that the Hamiltonian of a JJ in Eq.~\eqref{e.JJ} is the same as that of a fictitious particle in a cosine potential; in the limit $E_J \ll E_c$ (or $u \ll1 $) the lowest band has a width $w$ given by 
  \begin{equation}
  w = \frac{16}{\sqrt{\pi}} \left ( 8 E_J^3 E_c \right )^{1/4} e^{-\sqrt{\frac{8E_J}{E_c}}} = \frac{16}{\sqrt{\pi}} ~E_J ~u^{1/4} e^{-\frac{8}{\sqrt{u}}}
  \label{e.w}
  \end{equation}
  associated with an effective tunneling amplitude $J_{\rm eff} = w/4$. 
  
The appearance of a localized QPS can be attributed to the tunneling of the variable $\phi$ between two minima of the cosine potential -- in fact it involves the appearance of a $\phi=\pi$ phase difference across a junction, namely the climbing of the variable $\phi$ to the top of the $E_J$ barrier between two minima. Hence it is reasonable to expect that the QPS nucleation probability $p$ follows a similar functional dependence on $u$ as in Eq.~\eqref{e.w}.
 A similar result can be obtained within a Wentzel-Kramers-Brillouin approximation \cite{Sakurai} to the tunneling problem: as $E_c\sim U$ plays the role of the inverse mass of the fictitious particle with position $\phi$, the amplitude of the wavefunction within the barrier scales as $\exp(-\sqrt{m/m_0}) \sim \exp(-\sqrt{u_0/u}) $.
This exponential dependence of QPS probability on $u$ is clearly exhibited in Fig.~\ref{f.W2}(b), showing that for $u\leq 1$ the winding-number fluctuations follow the behavior $\langle W^2 \rangle \sim \exp(-b/\sqrt{u})$ with $b$ a constant. The essential singularity displayed by the quantum winding-number fluctuations as a function of $u$ in the limit of weak interactions clearly shows that this effect is due to highly non-linear quantum fluctuations, and it cannot be captured by a harmonic treatment such as Bogoliubov theory or Luttinger-liquid theory.  
 
  \begin{figure}[htb!]
 \centering
 \includegraphics[width=\columnwidth]{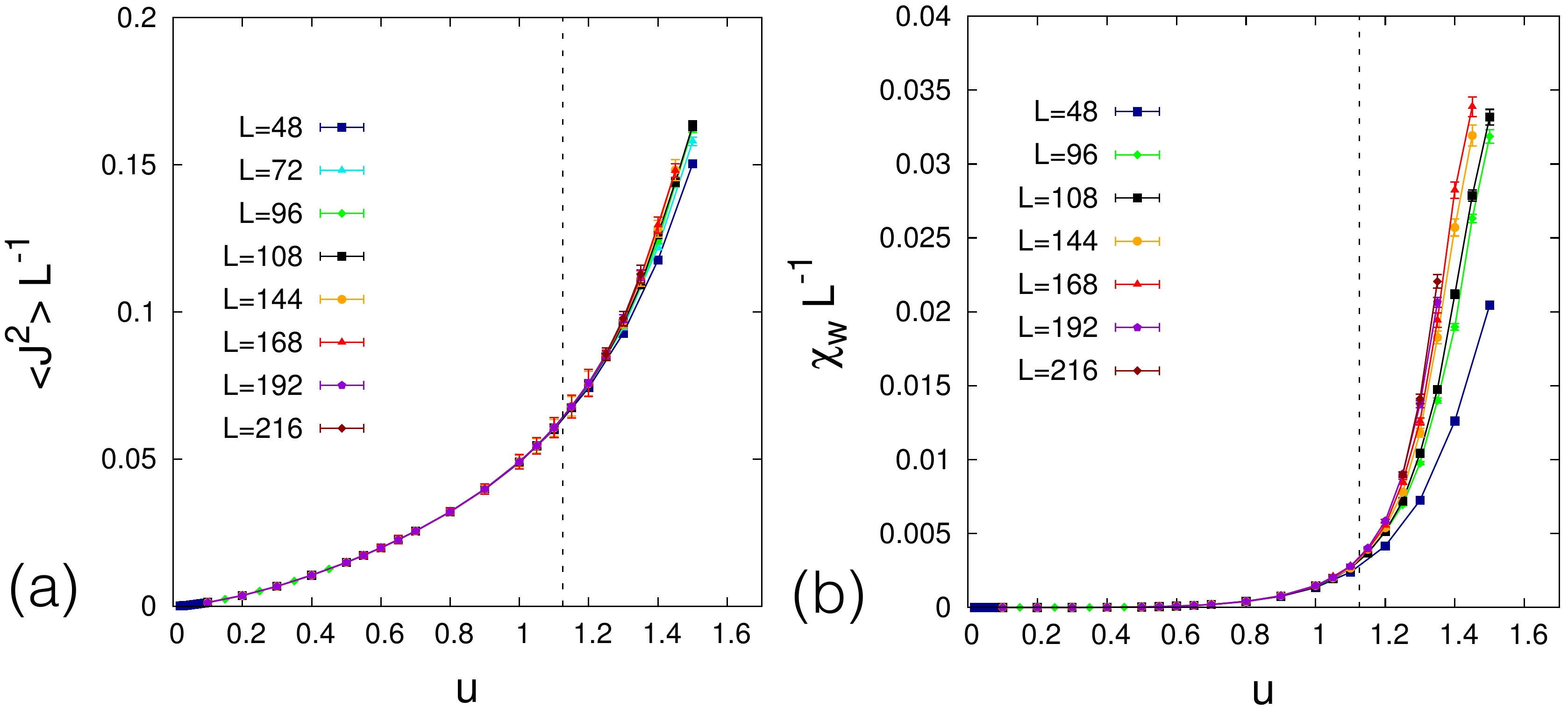}
\caption{Winding number susceptibility at $T=0$. The vertical dashed line marks the SF/MI transition.}
\label{f.J2W2}
\end{figure}

 \subsubsection{Quantum phase transition and Mott insulator phase}
As shown in Fig.~\ref{f.W2}(a), upon crossing the SF-MI transition one observes a strong enhancement of the winding number fluctuations, accompanied by a breakdown of the simple linear scaling with system size (Eq.~\eqref{e.quantumregime}). A similar behavior is also to be found in the  (equal-time) current fluctuations (see Fig.~\ref{f.J2W2}(a)):
  \begin{equation}
    \langle J^2 \rangle =  \left \langle \left [ \sum_i \sin(\phi_i - \phi_{i+1})  \right ]^2 \right \rangle =  \frac{1}{M} \sum_{k} \left \langle \left [ \sum_{i}  \sin(\phi_{i,k}-\phi_{i+1,k}) \right ]^2 \right \rangle_{\rm MC}~.   
    \label{e.ET}
 \end{equation}

 Our data do not allow us to conclude about a singular behavior (such as a jump) of $\langle W^2 \rangle$ or $\langle J^2 \rangle$ across the transition. As a matter of fact,  a numerical study of the thermal KT transition in the uniform 2$d$ XY model does not show any jump singularity of the corresponding observables, namely the winding number fluctuations or the current fluctuations along 1$d$ cuts \cite{Roscilde_unpublished}. On the other hand the a critical jump is expected for the current susceptibility $\chi_J$ of Eq.~\eqref{e.TA}, given that the jump in the  superfluid density $\rho_s$ in Eq.~(\ref{e.rhos}) at the SF/MI transition comes indeed from the current fluctuations and not from the Josephson term. Nonetheless, as is well known \cite{WeberM1988} and clearly documented by Fig.~\ref{f.SFMI}(b)), the numerical reconstruction of the jump is strongly limited by finite-size effects. Indeed one expects that \cite{WeberM1988}
 \begin{equation}
 \rho_s(u>u_c) \approx A ~e^{-a L/\xi(u)}  ~~~~ \to ~~~~ \chi_{J}(L) = L \left [ e(L) - A ~\frac{e^{-a L/\xi(u)}}{L}  + ... \right ]
\label{e.scaling}
\end{equation}
where $\xi(u)$ is the correlation length -- exponentially divergent as $u\to u_c^+$ -- and $a$ and $A$ are constants. The exponentially decreasing term in Eq.~\eqref{e.scaling} introduces the first correction to linear scaling (which becomes logarithmic exactly at the transition \cite{WeberM1988}, where $\xi(u) = \infty$), and its negative sign justifies the fact that the convergence of current fluctuations in the thermodynamic limit is attained from below. 

In the context of topological sector fluctuations, the counterpart to the current susceptibility is represented by the winding susceptibility
\begin{equation}
\chi_W = \int_0^{\beta} d\tau ~\langle W(\tau) W(0) \rangle = \frac{1}{t M^2} \sum_{kk'} \langle W_k W_k' \rangle_{\rm MC}~. 
\end{equation}
Fig.~\ref{f.J2W2}(b) shows that sizable deviations from linear scaling build up in $\chi_W$ at the transition, in a very similar manner to what happens to the current susceptibility. Therefore we conclude that a jump singularity is to be expected in the winding susceptibility, and that $\chi_W$ may obey a similar scaling as for $\chi_J$ in Eq.~\eqref{e.scaling}. Indeed strong winding-number correlations in imaginary time (whose integral gives $\chi_W$) imply strong correlations of the same kind for the current (while the opposite is not true, as current fluctuations may take place within the same topological sector). 

  \begin{figure}[htb!]
 \centering
 \includegraphics[width=0.7\columnwidth]{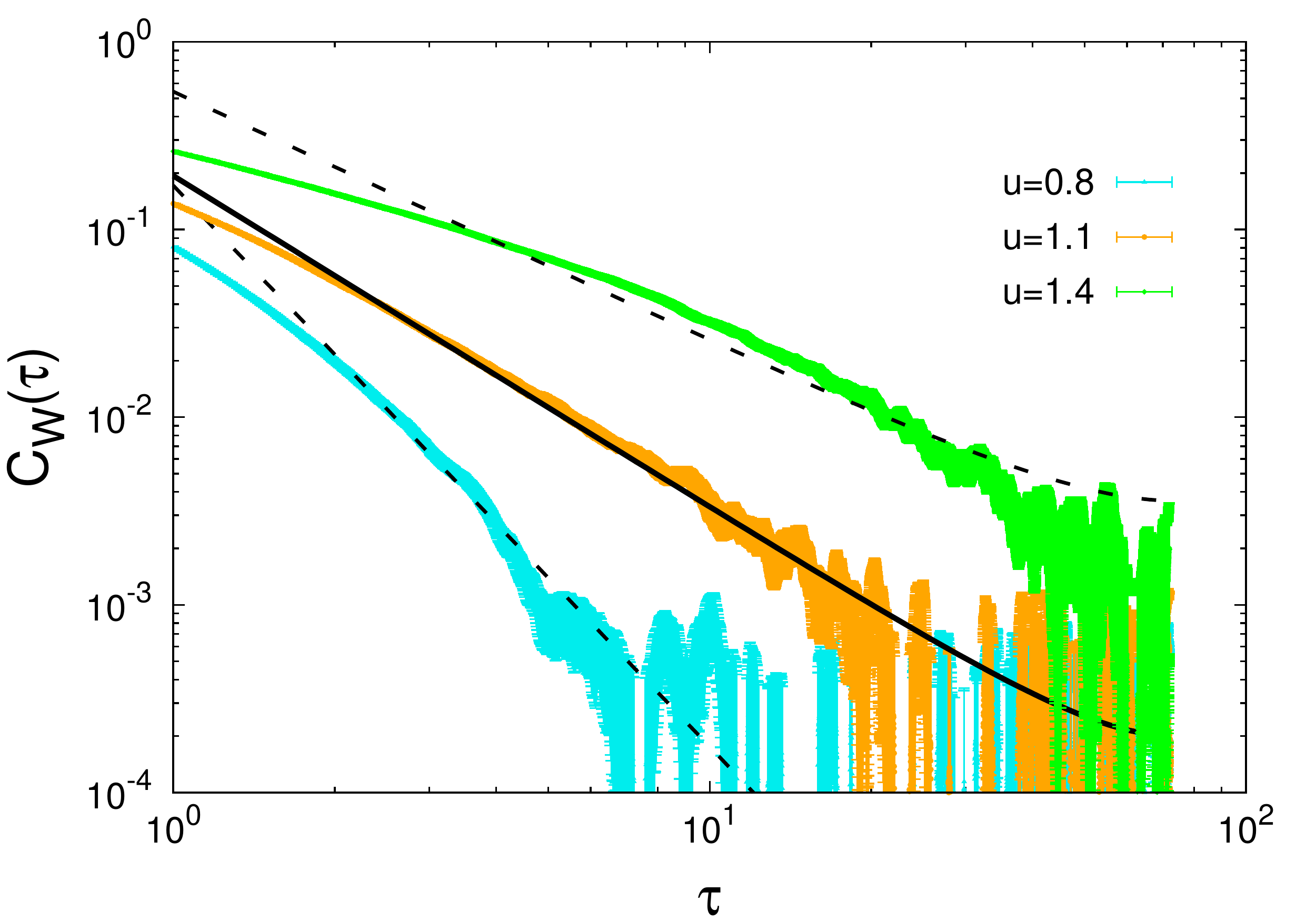}
\caption{Winding-number correlations in imaginary time for a $L=144$ chain at $t = 1/L$. Imaginary time $\tau$ on the $x$-axis is in units of $(2J{\bar n})^{-1}$. Solid and dashed lines are fits to $a[\tau^{-b} + (t^{-1}-\tau)^{-b}]$.}
\label{f.Wtau}
\end{figure}

It is tempting to associate the build-up of strong winding-number correlations in imaginary time with the unbinding of V-AV pairs along the same direction, leading to correlated winding-number fluctuations. Fig.~\ref{f.Wtau} shows indeed that close to the critical point $u_c$ the normalized winding number correlations 
\begin{equation}
C_W(\tau) =  \frac{\langle W(\tau)W(0) \rangle}{\langle W^2 \rangle}
\end{equation}
appear to develop a tail that is consistent with a power law, unlike in the MI phase, and with an exponent significantly smaller than that which can be estimated in the SF phase.  Our numerical results do not allow us to draw firm conclusions on the behavior of $C_W$ in the thermodynamic limit. Yet one may conjecture that, in the that limit and at $T=0$, $C_W$ decays as a fast power law deep in the SF phase (in the absence of any gap in the spectrum); and that this power-law decay changes abruptly at the critical point, with the further appearance of an exponential decay related to the opening of the Mott gap. All these decay laws are expected to be integrable -- given that the winding susceptibility keeps scaling linearly with $L$ across the transition, as imposed by the linear increase of $\langle W^2 \rangle$ -- but their integral leads to strong corrections to scaling as in Eq.~\eqref{e.scaling} when entering the MI phase. 
  
The standard picture of the KT transition associates the unbinding of V-AV pairs with the \emph{suppression} of conventional correlations in real space -- and the loss of coherence and stiffness as in Fig.~\ref{f.SFMI} is certainly a striking consequence of this aspect. On the other hand,  when focusing on winding-number fluctuations, the unbinding of V-AV actually leads to a critical enhancement of correlations for topological-sector fluctuations, namely ``topological correlations" are affected by topological defects in a rather different way from ordinary correlations. 
This point of view endows the jump in the superfluid stiffness in particular, and the 1$d$ SF/MI transition in general, with a precise topological nature, related to the singular behavior of winding-number correlations in imaginary time. The topological characterization of the SF/MI transition in the context of a 1$d$ lattice Bose gas parallels the one recently provided for the thermal KT transition of the 2$d$ lattice Coulomb gas in Ref.~\cite{Faulkneretal2015}; in the latter model, topological sectors are defined in terms of two winding numbers  (one per dimension), each of which map to the structures in Fig.~\ref{f.PI}(a). The transition is accompanied by a total suppression of these classical topological-sector fluctuations in the low-temperature phase.
  

 \begin{figure}[htb!]
 \centering
\includegraphics[width=\columnwidth]{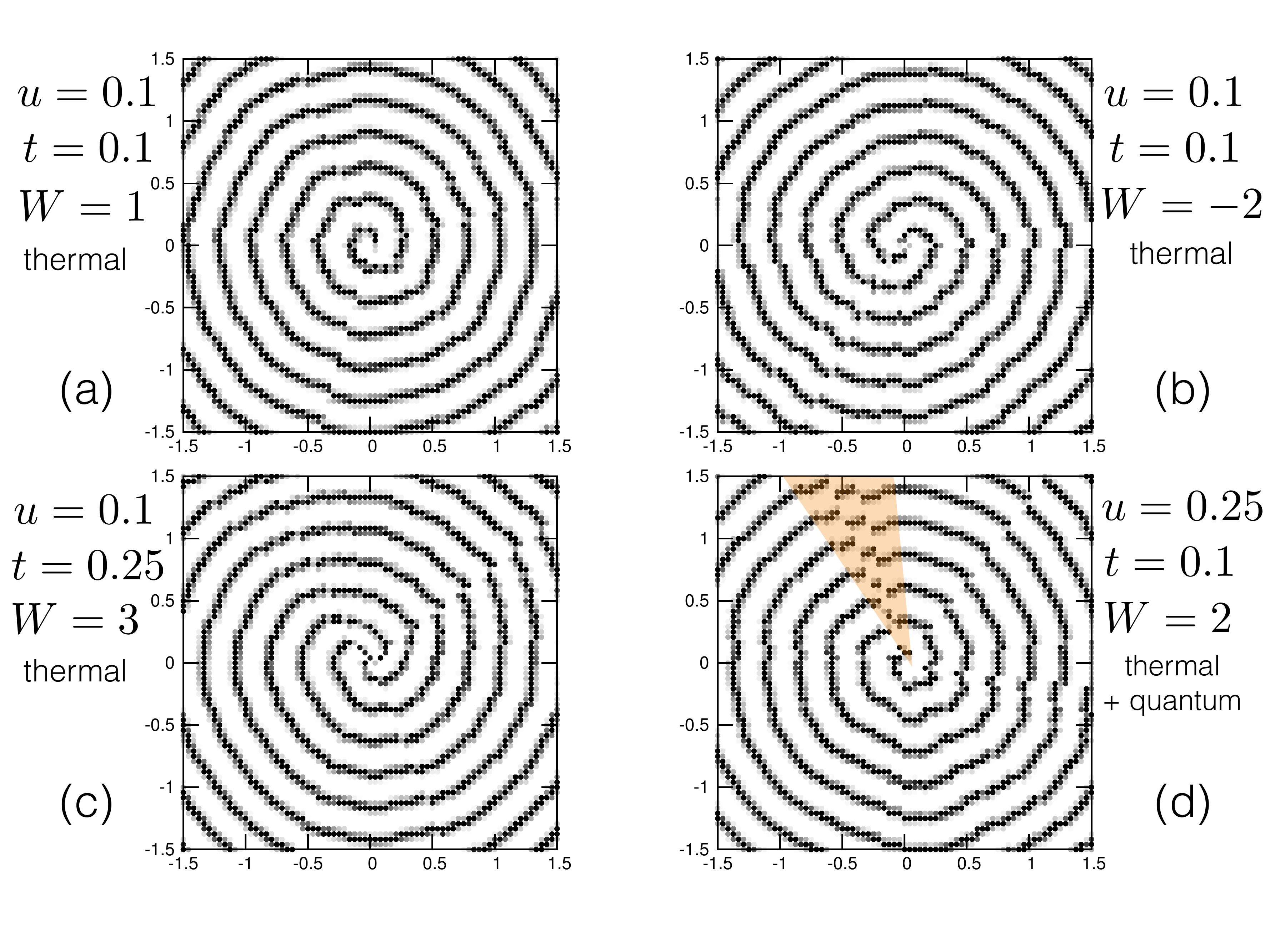}
 \caption{Inteference patterns in the presence of thermal phase slips (TPS) uniformly distributed along the ring, resulting from snapshots of the QMC simulation on a ring with L=48 and various temperatures $t$ and interaction strengths $u$.  The configurations have $|W|=$1, 2 and 3 (panels a-c), induced by thermal fluctuations; panel (d) shows a $W=2$ configuration with simultaneous presence of a uniformly distributed TPS and a localized QPS (marked by the shaded area) summing up.  The plotted function and color scale are the same as in Fig.~\ref{f.QPS}.}
\label{f.TPS}
\end{figure}

\section{Thermal regime: Arrhenius onset of smooth phase slips}
\label{s.thermal}

 When turning on the temperature, winding-number fluctuations can appear due to thermal excitations. Focusing our attention on the superfluid phase, we have seen in the previous section that phase configurations with $W=0$ largely dominate the ground state, and they are weakly admixed with $|W|\neq0$ configurations. The energy cost to create a finite $W$ for most of the configurations building up the ground state (or on most of the imaginary-time slices within the path-integral representation in Sec.~\ref{s.PI}) can be easily estimated via the superfluid stiffness defined in Eq.~\eqref{e.rhos}. Indeed the least energetic configuration with $W=\pm 1$ corresponds to that minimizing the energy in the presence of twisted boundary conditions, which is achieved by distributing the $\pm 2\pi$ phase twist uniformly along the ring, namely by developing a phase difference $\varphi = \pm 2\pi/L$ across each junction in the lattice. In the classical limit of the quantum phase model ($U=0$) the energy cost for such a configuration would be simply 
 \begin{equation}
\frac{\Delta E_{|W|=1} }{2J\bar{n}} =  L[\cos(2\pi/L)-1] \approx \frac{2\pi^2}{L}
 \end{equation}
 where $\Delta E_{|W|=1} = E(|W|=1) -E(0)$ and we consider the limit $L\gg 1$. 
 Quantum fluctuations induced by $U$, namely quantum phonons and QPS, renormalize this energy cost, which is properly estimated via the zero-temperature superfluid fraction   
 \begin{equation}
 \rho_s(t=0) \approx \frac{1}{2J{\bar n}L} \frac{E(\varphi) + E(-\varphi) - 2E(0)}{\varphi^2} =  \frac{L}{2J{\bar n}} \frac{\Delta E(\varphi)}{2\pi^2} ~~~\to~~~\frac{\Delta E_{|W|=1}}{2J{\bar n}} \approx\frac{2\pi^2\rho_s}{L}~.
 \end{equation}
 Similarly the energy cost for a generic winding number $W$ can be estimated as  $\Delta E_{|W|}/(2J{\bar n}) \approx 2\pi^2\rho_s W^2/L$, provided that $L \gg 2\pi W$. 
According to this estimate, one expects that thermal fluctuations produce configurations possessing uniformly distributed $2\pi W$ phase twists with an Arrhenius probability
$ P(W) \sim \exp[-2\pi^2 W^2 \rho_s/ (tL)] $, leading to winding number fluctuations with variance
\begin{equation}
\langle W^2 \rangle \sim A \exp[-2\pi^2 \rho_s/(tL)] + 4 B \exp[-8\pi^2  \rho_s/(tL)] + ...  
\label{e.W2therm}
\end{equation}
where $A, B, ...$ are multiplicity factors. 
Fig.~\ref{f.TPS}(a-c) shows interference patterns reconstructed from QMC snapshots bearing a finite winding number $W$ (up to $W=3$) in the regime of thermal excitation of low-energy TPS on top of a superfluid ground state. Such low-energy TPS result in very smooth spirals in the interference pattern, decorated with localized dislocations of the fringe pattern due to quantum fluctuations. Moreover, for sufficiently large $u$ one often observes configurations exhibiting both a TPS and a QPS, clearly distinguished by the extended nature of the former and the localized nature of the latter (see Fig.~\ref{f.TPS}(d) for an example, as Sec.~\ref{s.crossover} for further discussion).  

\begin{figure}[htb!]
 \centering
\includegraphics[width=0.8\columnwidth]{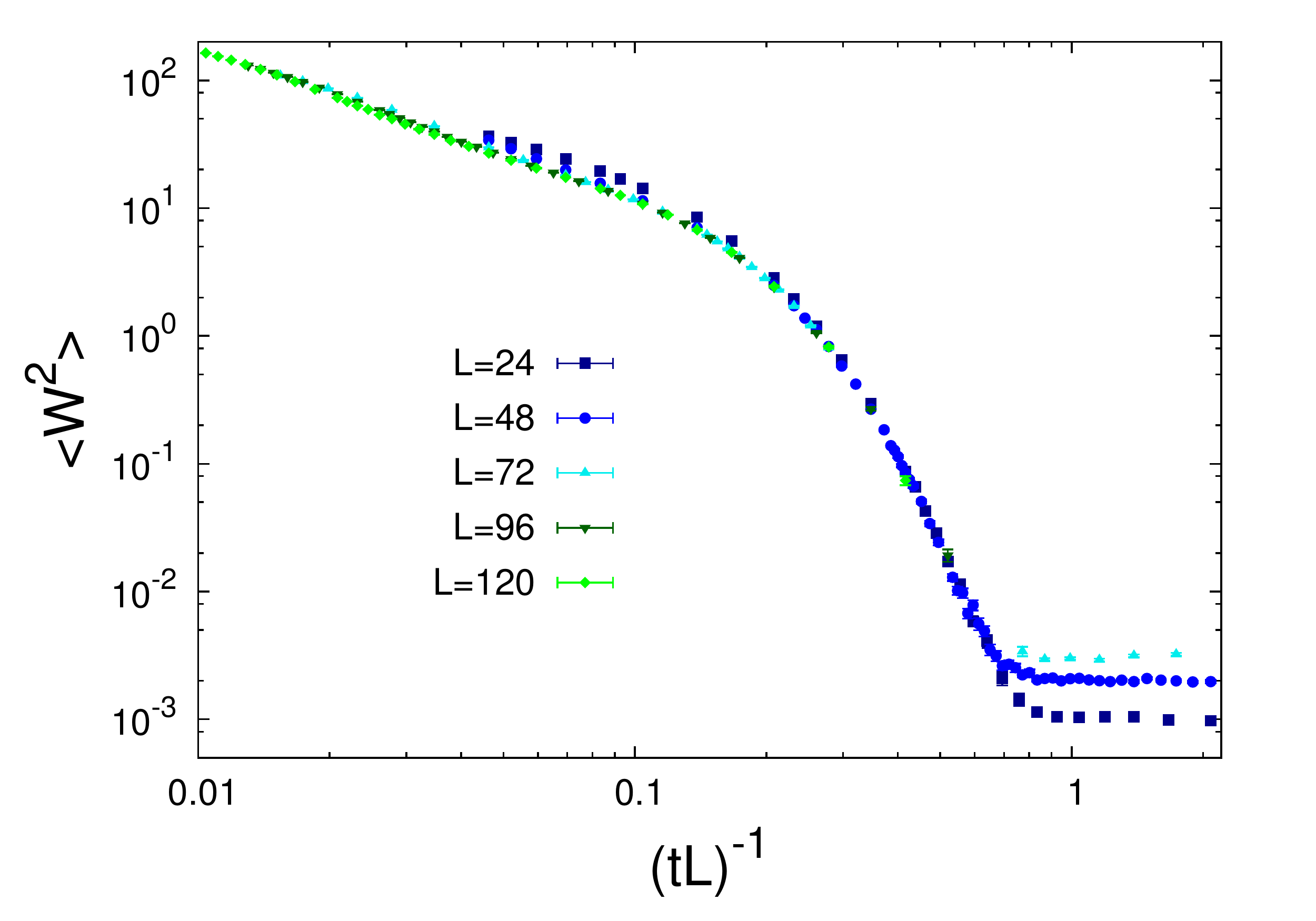}
 \caption{Scaling of thermal winding number fluctuations for $u=0.25$, showing consistency with the Ansatz of Eq.~\eqref{e.thermalregime} over a broad temperature range.}
\label{f.W2thermscaling}
\end{figure}

The Arrhenius law for winding number fluctuations in Eq.~\eqref{e.W2therm} immediately implies an exponential sensitivity to the system size, and the general scaling behavior 
\begin{equation}
\langle W^2 \rangle(T;L) = F_{W}(T\times L)
\label{e.thermalregime}
\end{equation}
which is obviously inconsistent with the scaling behavior of Eq.~\eqref{e.quantumregime} valid in the quantum regime. As shown in Fig.~\ref{f.W2thermscaling} such a scaling form is indeed very well verified by our data down to a characteristic crossover temperature at which the thermal regime breaks down, leaving space to the quantum regime which rather obeys the Poissonian scaling as in Eq.~\eqref{e.quantumregime}. The detailed study of the crossover will be the subject of Sec.~\ref{s.crossover}. 

\begin{figure}[htb!]
 \centering
\includegraphics[width=\columnwidth]{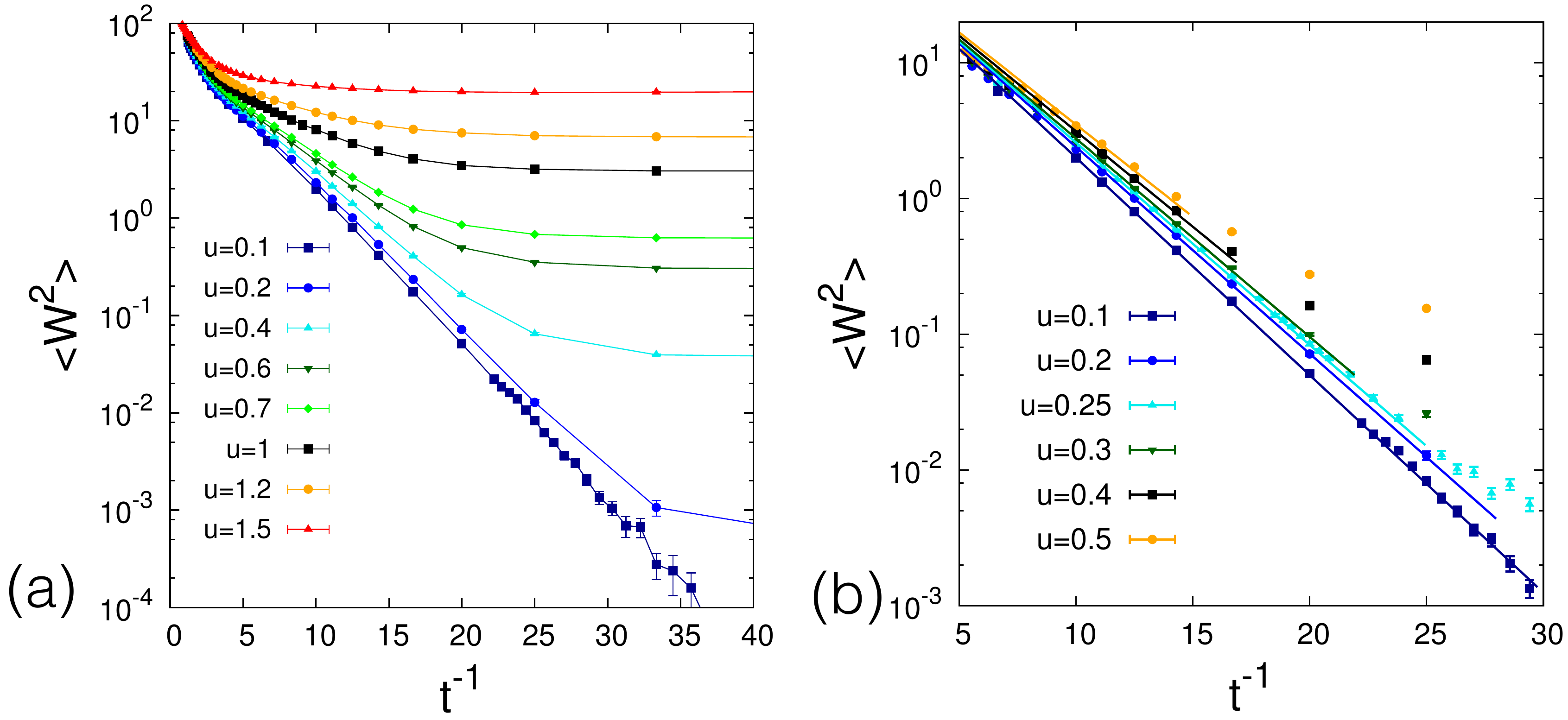}
 \caption{(a) Thermal onset of winding-number fluctuations for a chain of $L=48$ sites and variable interaction strengths; (a) same as in panel (b), but focusing uniquely on the weakly interacting regime; solid lines are functions $A\exp[-2\pi^2\rho_s(u,t=0)/(tL)]$, where $\rho_s(u,t=0)$ is the superfluid fraction plotted in Fig.~\ref{f.SFMI}(b), and $A\approx80$.}
\label{f.W2therm}
\end{figure}

 Moreover Eq.~\eqref{e.W2therm} implies an exponential temperature dependence of the thermal winding-number fluctuations, dominated by the $|W|=1$ term. This exponential dependence is indeed clearly observed in our data within the SF regime ($u\lesssim u_c$), as shown in Fig.~\ref{f.W2therm}(a): in particular when $t\to 0$ the exponential decrease of $\langle W^2 \rangle$ dominates the intermediate temperature range $t \lesssim 1$, but it then levels off at a $u$- (and size-) dependent temperature, marking the crossover to the regime of quantum winding-number fluctuations, insensitive to temperature. As predicted by Eq.~\eqref{e.W2therm} the exponential decay rate of $\langle W^2 \rangle$ is controlled by the zero-temperature superfluid density, as it is very well verified by the data in Fig.~\ref{f.W2therm}(b). This in turn shows that the study of the thermal dependence of the winding-number fluctuations is an exponentially sensitive probe of the superfluid density; we will come back to this point in the Conclusions.

\begin{figure}[htb!]
 \centering
\includegraphics[width=0.8\columnwidth]{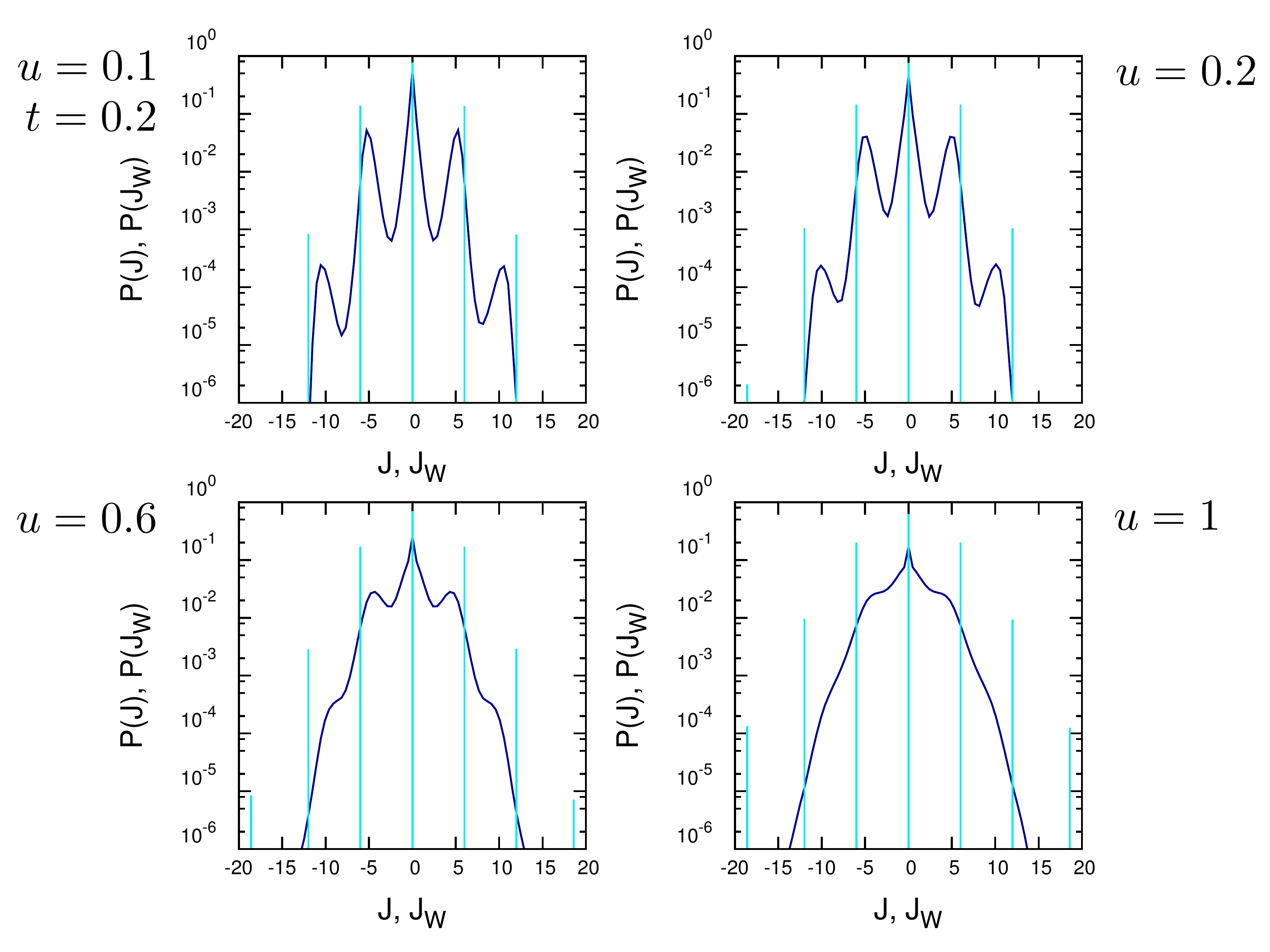}
 \caption{Probability distribution functions for current fluctuations on a $L=48$ ring at $t=0.2$ and variable interactions $u$. The dark blue line is $P(J)$, while the cyan line is $P(J_W)$, descending from the probability distribution for winding number fluctuations.}
\label{f.histoJ}
\end{figure}

 So far we have identified the thermal regime of winding-number fluctuations with: 1) the peculiar scaling form it satisfied by fluctuations as in Eq.~\eqref{e.thermalregime}; 2) the exponential activation of fluctuations controlled by the (zero-temperature) superfluid fraction; and 3) by the uniformly twisted phase configurations that are thermally populated in the finite-$W$ sector at sufficiently low temperatures. This latter aspect reflects itself not only in the qualitative aspects of the experimentally accessible interference patterns (as shown in Fig.~\ref{f.TPS}), but also (and more quantitatively) in a peculiar structure for the  distribution of current fluctuations, which are accessible experimentally via the interferometric reconstruction of the phase differences $\phi_i-\phi_{i+1}$ across the tunnel barriers.  Indeed a smooth phase twist with winding number $W$ displays phase differences $\phi_i-\phi_{i+1}\approx 2\pi W/L$ among neighboring sites, leading to a current $J\approx J_W = L \sin(2\pi W/L)$. If one limits oneself to smooth phase twists only, current fluctuations should develop a discrete distribution with peaks corresponding to the discrete $W$ values. Such a picture is indeed realized in the regime of weak interactions $u \ll 1$ and low temperatures, where non-smooth twists are prevented from forming both quantum-mechanically and thermally -- see Fig.~\ref{f.histoJ}; on the other hand, upon increasing $u$ the multipeak structure is rounded off when the occurrence of localized QPS equals and eventually overcomes that of the smooth TPS, leading to unconstrained current fluctuations which progressively develop into a Gaussian distribution. 

\begin{figure}[htb!]
 \centering
\includegraphics[width=0.7\columnwidth]{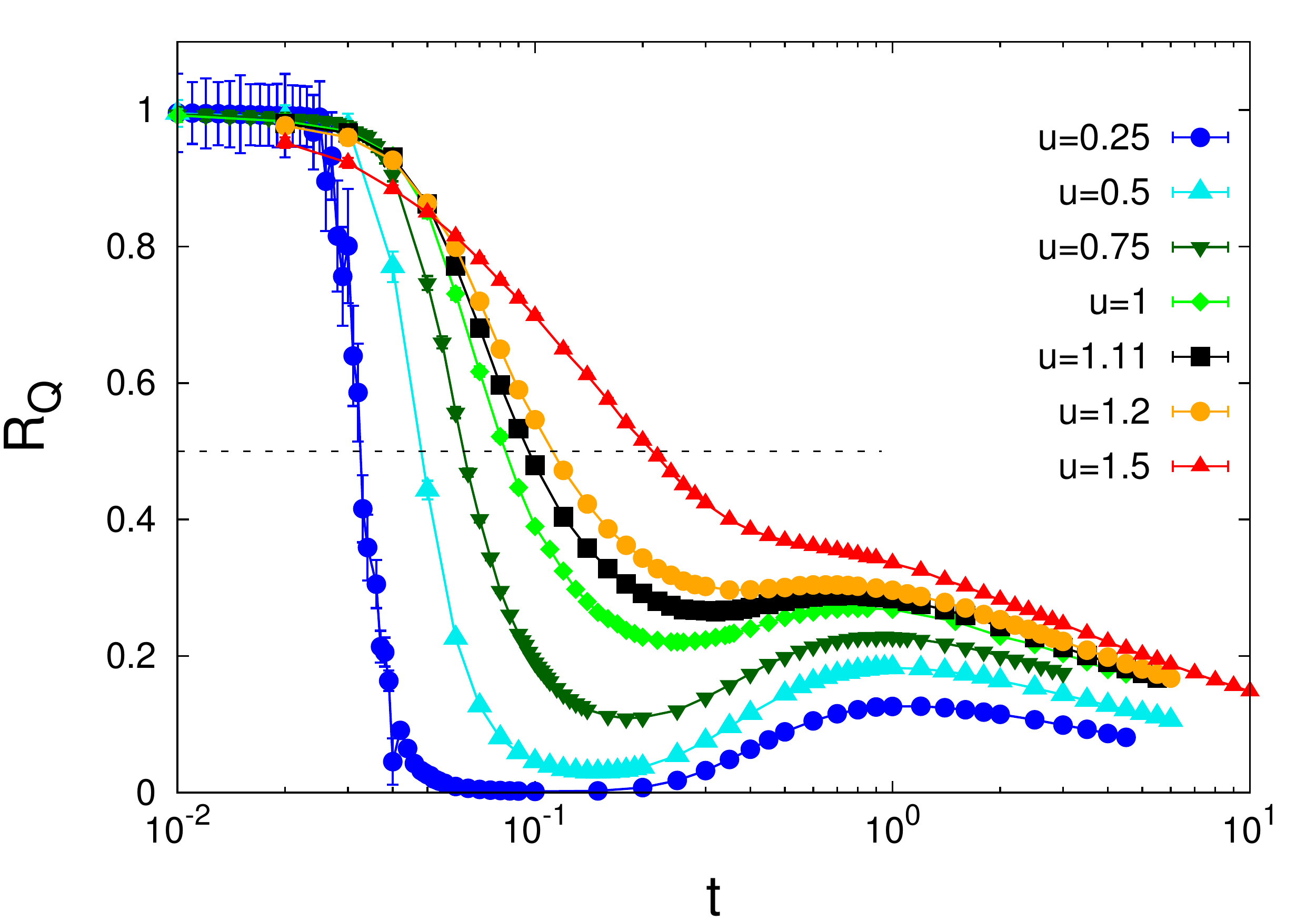}
 \caption{Ratio $R_Q$ between quantum coherent and total winding-number fluctuations as a function of temperature for a ring with $L=48$ and different values of the interaction strength. The dashed line marks the thermal-quantum crossover $R_Q = 1/2$.}
\label{f.RQ}
\end{figure}

\section{Quantum-to-thermal crossover and thermal revival of quantum phase slips}
\label{s.crossover}

 \subsection{Quantum-to-thermal crossover, and thermal revival of quantum fluctuations, from the quantum-to-total fluctuation ratio}

 As pointed out in Secs.~\ref{s.quantum} and \ref{s.thermal}, thermal and quantum winding-number fluctuations display a rather different scaling behavior, and they possess as well rather different spatial properties. Hence a most interesting aspect is to locate the crossover between the two regimes, and to determine its most important signatures. To achieve the first goal unambiguously, we can rely again on the path-integral formulation of the partition function for the quantum phase model discussed in Sec.~\ref{s.PI}, identifying quantum-coherent winding-number fluctuations with those taking place along the imaginary-time dimension. In contrast, thermal (or incoherent) fluctuations do not involve the imaginary-time direction, but they originate from fluctuations between different path-integral configurations contributing to the partition function. In order to eliminate the information on the imaginary-time fluctuations one can define the imaginary-time-averaged winding number
 \begin{equation} 
 \bar{W} = \frac{1}{\beta} \int_0^{\beta} d\tau ~W(\tau) = \frac{1}{M} \sum_k W_k
 \end{equation} 
 and hence define the thermal winding-number fluctuations as those of $\bar W$:
 \begin{equation}
 \langle W^2 \rangle_T = t ~\chi_W = \langle \bar{W}^2 \rangle = \frac{1}{M^2} \sum_{k,k'} \langle W_k W_{k'} \rangle_{\rm MC} 
  \end{equation}
 It is easy to prove that $\langle W^2 \rangle_T \leq \langle W^2 \rangle$, namely the total fluctuations always exceed the thermal ones, because of the very existence of imaginary-time fluctuations. Hence quantum winding number fluctuations can be estimated as the residual fluctuations \cite{FrerotR2015}:
 \begin{equation}
 \langle W^2 \rangle_Q = \langle W^2 \rangle -  \langle W^2 \rangle_T~.
 \label{e.W2Q}
 \end{equation}
 In the ground state, namely in the limit in which the imaginary-time direction becomes infinitely long, each path-integral configuration should sample the whole statistics of fluctuations associated with the quantum ground state $|\psi_0\rangle$; this implies that $\bar{W} = \langle \psi_0 | W | \psi_0 \rangle = 0$ due to time-reversal symmetry. As a consequence the configuration-to-configuration fluctuations $\bar{W}$ vanish, and, as expected, quantum fluctuations as in Eq.~\eqref{e.W2Q} saturate the total fluctuations. The fluctuations of $\bar{W}$ are instead thermally activated; as a consequence, we can evaluate the relative strength of quantum fluctuations with the ratio 
\begin{equation}
 R_Q = \frac{\langle W^2 \rangle_Q}{\langle W^2 \rangle}
\end{equation} 
satisfying the two limiting behaviors $R_Q \to 1$ for $t \to 0$ and $R_Q \to 0$ for $t\to \infty$. It is then natural to define a (size- and parameter-dependent) \emph{crossover temperature} $t_x$ between the quantum and thermal regime of winding number fluctuations as the one satisfying the criterion:
\begin{equation}
R_Q(t_x) = 1/2~.
\label{e.RQ}
\end{equation}
Fig.~\ref{f.RQ} shows $R_Q$ as a function of $t$ for different values of the interaction strength: the crossover temperature systematically moves to higher values as $u$ increases,  as expected due to the increase of the strength of quantum winding-number fluctuations.

\begin{figure}[htb!]
 \centering
\includegraphics[width=0.7\columnwidth]{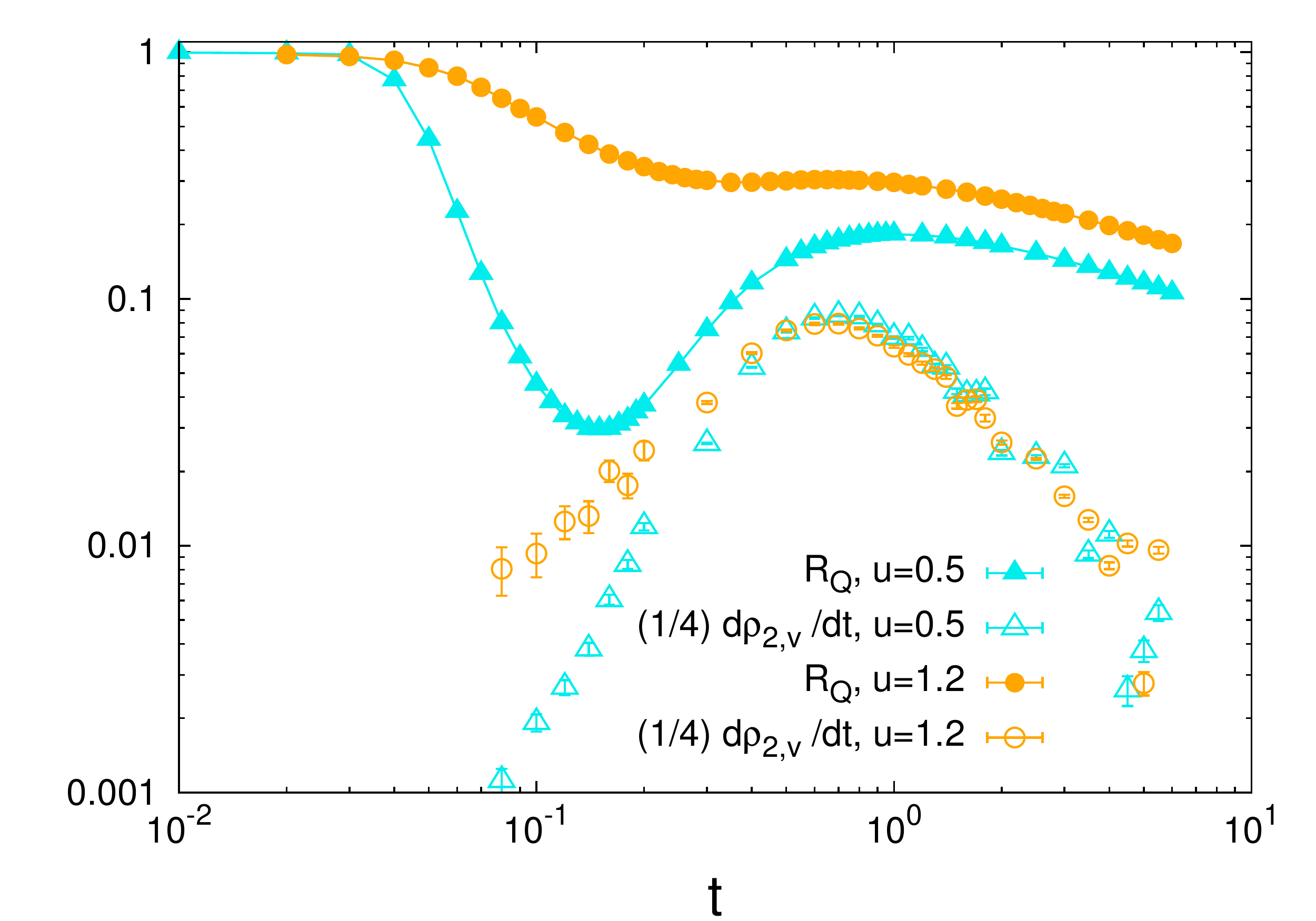}
 \caption{Comparison between $R_Q$ ratio and the temperature derivative of vorticity $d\rho_{2,v}/dt$ for a ring with $L=48$ and $\epsilon = 10^{-2}$.}
\label{f.RQvortex}
\end{figure}

\begin{figure}[htb!]
 \centering
\includegraphics[width=\columnwidth]{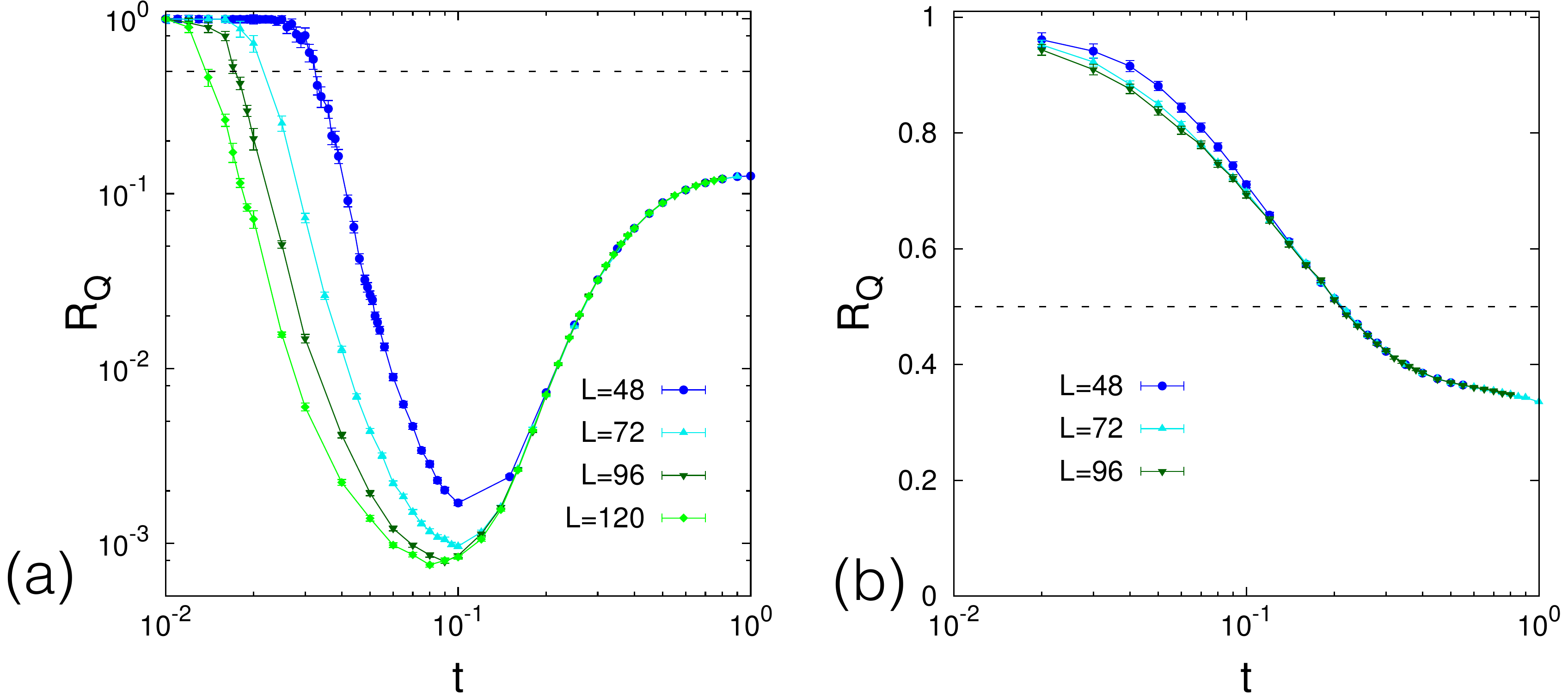}
 \caption{Scaling of the $R_Q$ ratio for (a) $u=0.25$ (SF phase) and (b) $u = 1.5$ (MI phase). The dashed line marks the value 1/2.}
\label{f.RQscaling}
\end{figure}

 Beside the growth of the crossover temperature with interaction strength, one can observe that in the SF phase a highly non-trivial interplay between thermal and quantum winding-number fluctuations is exhibited by the $R_Q$ data when looking at a broad temperature scale. Indeed $R_Q$ is found to display a non-monotonic behavior in the SF phase, { with a further maximum at finite $T$, shifting to lower temperatures as $u$ increases}. This feature is found not to depend on the value of $\epsilon$ we chose (namely it is a feature of the quantum phase model as such), and, within the SF phase, it is also not dependent on the size, as shown in Fig.~\ref{f.RQscaling}. A possible interpretation of this peculiar thermal revival of quantum fluctuations stems from the mapping of the 1$d$ quantum phase model onto the anisotropic 2$d$ XY model of Eq.~\eqref{e.XY}, suggesting that quantum winding-number fluctuations are induced by the appearance of V-AV excitations separating along the imaginary-time dimension (Fig.~\ref{f.PI}(b)). 
{ As we have seen in Sec.~\ref{s.XYmapping},  at fixed $\epsilon$ all couplings constants are held fixed and the temperature uniquely controls the length of the extra dimension, $M = u/(t\epsilon)$}. The thermal shrinking of $M$ drives the system from fully 2$d$ at $T=0$ to quasi-1$d$ at finite temperature, and this dimensional crossover alone is found to be sufficient to enhance the vorticity in the system. 
We can quantitatively relate the thermal enhancement of quantum winding-number fluctuations to the proliferation of V-AV pairs by calculating the squared vortex density of the fictitious 2$d $XY model of Eq.~\eqref{e.XY}
\begin{equation}
\rho_{2,v} = \frac{1}{ML}  \sum_{\square} \left\langle \left ( \sum_{i\to j \in \square} F(\phi_j-\phi_i) \right )^2 \right \rangle_{\rm MC}
\end{equation}
where $F$ is the function defined in Eq.~\eqref{e.F}, $\sum_{\square}$ runs over the space-time square plaquettes and $\sum_{i\to j}$ runs counterclockwise over the four sites of each
plaquette. This quantity is an exclusive property of the path-integral representation in discrete imaginary time, and it does \emph{not} correspond to a physical observable of the quantum-phase model (in fact, it is found to depend on the choice of $\epsilon$). Yet, as shown in Fig.~\ref{f.RQvortex}, its temperature derivative $d\rho_{2,v}/ dt$ exhibits a maximum which correlates very well with the finite-$T$ maximum of the $R_Q$, slightly shifting to lower temperatures when the interaction increases. The temperature dependence of $\rho_{2,v}$ stems entirely from the dimensionality crossover described above: as the system moves from a torus geometry to an annulus geometry upon decreasing M, its stiffness (in the sense of the XY model  of Eq.~\eqref{e.XY2}) opposing to the thermal activation of V-AV pairs is gradually reduced -- given that, all other parameters being held fixed, the stiffness of an XY model is higher in $d=2$ than in $d=1$.

 \subsection{Crossover from measurable quantities}

 The quantum-to-thermal crossover in winding number fluctuations at low temperature, and the unusual phenomenon of thermal enhancement of quantum fluctuations at intermediate temperature, is not only revealed theoretically by the $R_Q$ ratio, but it is also witnessed directly by measurable quantities. In the Appendix (Sec.~\ref{s.app1}) we show representative interference patterns obtained from uncorrelated snapshots of the MC simulation, mimicking the outcomes of uncorrelated shots of the experiment. In contrast to that found at $T=0$ (see Fig.~\ref{f.QPS}) we observe that increasing the temperature above zero first leads to the appearance of uniform phase slips leading to smooth spirals in the interference patterns, corresponding to TPS which fully dominate the path-integral configuration at all imaginary-time slices; only rarely does one observe localized phase slips (both in space and in imaginary time), corresponding to quantum fluctuations -- this is consistent with the initial drop of $R_Q$. On the other hand, as the temperature grows further the localized phase slips are seen to be increasingly frequent, in accordance with the thermal increase of $R_Q$. 
   
   We can make this qualitative observation on the geometry of the interference fringes much more quantitative by defining the distance of finite-$W$ phase configurations at fixed imaginary time, $\{\phi_{i,k}\}_{i=1,...,N}$, from a perfectly uniform phase slip with the same $W$, namely  
  \begin{equation} 
   \Delta_W = \left \langle \sqrt{ \min_i \sum_{j=1}^L \left [ \phi_{i+j,k}-\mod(\phi_{i,k} + 2\pi W j/L,2\pi) \right ]^2  ~} \right \rangle_{W}. 
  \end{equation} 
  where $\langle ... \rangle_{W}$ represents the thermal average restricted to the topological sector with winding number $W$.  
  The above quantity looks for the minimal distance between the phase configuration and a configuration exhibiting a uniform phase twist of $2\pi W$, starting from any site $i$ on the lattice (here $i+j$ is defined modulo $L$).  In the MC simulation we then average the distance over all configurations exhibiting the same finite winding number $|W|$ (in absolute value). 
  
  Another sensitive observable capturing the quantum-to-thermal crossover is given by the integrated current correlations at finite winding-number  
  \begin{equation}
  S_{J,W} = \frac{1}{L} \left \langle \sum_{ij} \sin(\phi_{i,k}-\phi_{i+1,k}) \sin(\phi_{j,k}-\phi_{j+1,k}) \right\rangle_W
 \end{equation}
 restricted to the topological sector with winding number $W$. A uniform phase slip with winding number $W$ exhibits strong current-current correlations, $S_{J,W} = L \sin^2(2\pi W/L) \approx (2\pi W)^2/L$; on the other hand, such correlations are suppressed if the phase slip is localized: in the limit of a phase slip localized \emph{e.g.} on three bonds (with a slip of $2\pi W/3$ each), $S_{J,W} \approx 9 \sin^2(2\pi W/3)/L$. For $W=1$ and $L=48$, $S_{J,1} \approx 0.82$ in the first case, while $S_{J,1} \approx 0.14$ in the second case. 
 
\begin{figure}[htb!]
 \centering
\includegraphics[width=0.7\columnwidth]{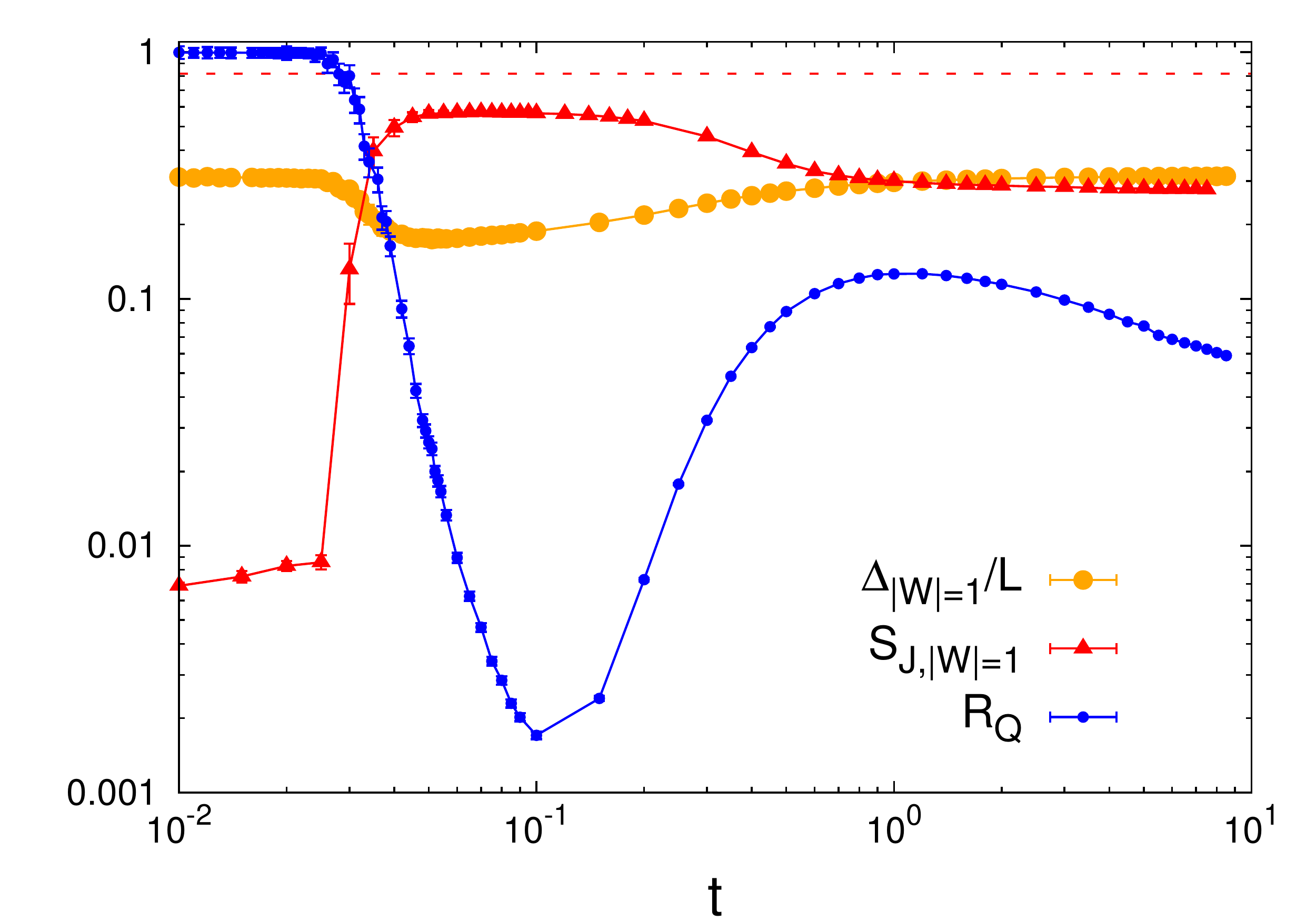}
 \caption{Distance to a uniform phase slip $\Delta_{|W|=1}$ and integrated current-current correlations at finite winding number $S_{J,|W|=1}$ for a $L=48$ ring with $u=0.25$; these two quantities are compared to the $R_Q$ ratio, already shown in Fig.~\ref{f.RQ}. The dashed line shows the maximum value $L \sin^2(2\pi W/L)$ expected for $S_{J,|W|=1}$.}
\label{f.DeltaSJ}
\end{figure}

 The average $\Delta_{|W|=1}$ and $S_{J,|W|=1}$ are plotted in Fig.~\ref{f.DeltaSJ}, showing that the quantum-to-thermal crossover at $t_x$ is marked by an abrupt drop of the distance to a uniform phase slip, accompanied by a sudden enhancement of current-current correlations. This shows that there is a fundamental correspondence between the coherent nature of the phase slips leading to winding-number fluctuations, and the spatial structure of such excitations, which is fully accessible to experiments. 
 Moreover, the two quantities under investigation show a broad minimum and a plateau (respectively) over the temperature range in which $R_Q$ nearly vanishes, marking the thermal regime. On the other hand, upon increasing the temperature $\Delta$ shows a revival, and $S_J$ a suppression, both corresponding to the revival in the $R_Q$. The enhancement of quantum winding-number fluctuations is certainly responsible for the temperature dependence of $\Delta$ and $S_J$, but one should add to this the contributions of thermally populated phonons which, while not affecting at all $R_Q$, are certainly responsible for a distortion of smooth spirals with respect to the low-$T$ regime, and for a weakening of current-current correlations.

\begin{figure}[htb!]
 \centering
\includegraphics[width=0.9\columnwidth]{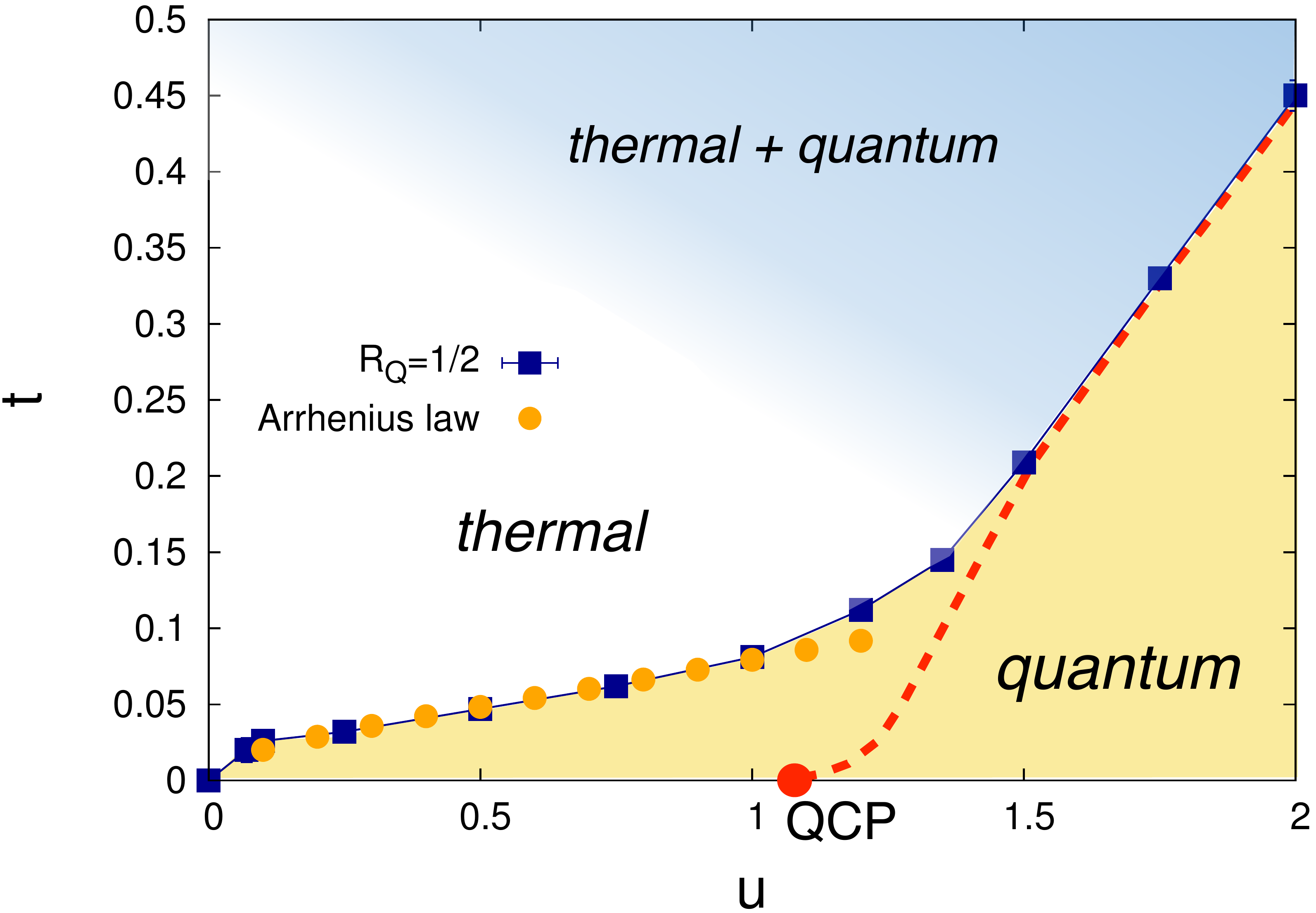}
 \caption{Quantum-to-thermal crossover in the winding-number fluctuations of a $L=48$ ring:  the criterion $R_Q=1/2$ (see text) is compared with the one of Eq.~\eqref{e.crossover} (stemming from Arrhenius law for thermal excitations). The dashed red line sketches the expected behavior in the thermodynamic limit.}
\label{f.crossover}
\end{figure}

 \subsection{Interaction- and size-dependence of the crossover temperature; phase diagram} 
 
 We conclude the discussion of the quantum-to-thermal crossover by focusing on its dependence on the system size and interaction strength. In the SF case such dependence can be readily extracted by identifying the crossover temperature with the one at which thermal winding-number fluctuations, setting in with the Arrhenius law of \eqref{e.W2therm}, overcome the $T=0$ quantum fluctuations, namely
\begin{equation}
A \exp\left(\frac{-2\pi^2\rho_s(t=0;u)}{L t_x}\right) \approx \langle W^2 \rangle(t=0;u)
\label{e.crossover}
\end{equation}
which, when solving for $t_x$, gives
\begin{equation}
t_x(u;L) \approx \frac{ 2\pi^2 \rho_s(t=0;u)}{L \left |\log\left(\frac{\langle W^2 \rangle_{t=0;u}}{A}\right)\right|}~.
\label{e.tx}
\end{equation}
Hence we observe that the crossover temperature vanishes as $1/L$ (up to logarithmic corrections -- see also Fig.~\ref{f.RQscaling}(a)), and that its interaction dependence stems from a competition between the interaction dependence of $\rho_s$ and $\langle W^2 \rangle(t=0;u)$. Indeed, while $\rho_s$ tends to vanish upon increasing $u$, the ground-state fluctuations $\langle W^2 \rangle$ increase above zero, eliminating the divergence of $\log \langle W^2 \rangle$ in the denominator. It turns out that the latter effect wins for weak interactions, given also the exponential onset of the quantum winding-number fluctuations following Eq.~\eqref{e.w} (see also Fig.~\ref{f.W2}(b)).

 Fig.~\ref{f.crossover} shows that, sufficiently deep in the SF phase, Eq.~\eqref{e.crossover} provides an excellent estimate of the finite-size crossover temperature when compared with the one satisfying the criterion of Eq.~\eqref{e.RQ}. 
Eq.~\eqref{e.tx} clearly shows that the quantum-to-thermal crossover in the SF regime is a mesoscopic effect, and that the crossover temperature is pushed to zero as $L\to\infty$ due to the collapse of the finite-$W$ topological sectors onto the $W=0$ sector which dominates the ground state { of} finite-size systems. The size dependence of the crossover temperature, along with the size dependence of quantum winding-number fluctuations (Eq.~\eqref{e.quantumregime}) points towards the existence of an optimal size range for the experimental observation: indeed, while a small size $L$ enhances the crossover temperature, the number of measurements required to observe Poissonian quantum winding-number fluctuations decreases with $L$.

On the contrary, in the MI regime the crossover temperature remains finite even in the thermodynamic limit, as shown in Fig.~\eqref{f.RQscaling}(b). The ground state already features a strong coherent admixture of different topological sectors, and ground state physics is protected by the Mott gap: winding-number fluctuations therefore maintain quantum coherence up to a temperature which should be comparable with the gap itself.  In Fig.~\ref{f.crossover} we sketch therefore the crossover temperature in the thermodynamic limit, vanishing at the SF/MI quantum critical point, and building up as $u\geq u_c$ in a manner similar to the Mott gap.

\section{Conclusions}
\label{s.conclusions}

 In this paper we have shown that winding number fluctuations in lattice Bose rings display a very rich behavior, exhibiting a quantum regime of localized quantum phase slips, and a thermal regime of extended thermal phase slips. The fundamental correspondence between the thermal or quantum nature of phase slips, and their spatial structure, makes this crossover detectable in state-of-the-art experiments on ultracold gases, which are able to fully reconstruct the phase configuration of the ring-trapped Bose gas via matter-wave interference.  It is also worth noticing that the winding-number fluctuations stand as a direct probe of the ground-state superfluid density, even without probing the response of the system to rotation or to a gauge field. Indeed, by looking at the thermal winding-number fluctuations one restricts naturally to topologically non-trivial sectors of the phase configurations, in which the system has effectively come into rotation spontaneously. The ground-state superfluid density is exactly what opposes the appearance of such fluctuations, and its renormalization due to quantum fluctuations is directly observable as a softening of thermal winding-number fluctuations. 
 
  Our work defines the experimental conditions for the direct observation of quantum phase slips in atom-circuit setups, such as the ones explored in Refs.~\cite{Cormanetal2014, Eckeletal2014}. This perspective nicely complements the recent effort in the context of superconducting nanostructures \cite{Popetal2010,Manucharyanetal2012,Astafievetal2012,Erguletal2013}, fully exploiting the power of atom interferometry. Further extensions of our work can be readily envisioned to involve topological excitations in higher-dimensional systems, or the study of quantum-quench dynamics across the superfluid-insulator transition.

\section{Ackowledgements}
Fruitful discussions with A. Ran\c con are gratefully acknowledged. This work is supported by ANR (`ArtiQ' project).

\section{APPENDIX: Interference patterns at the quantum-to-thermal crossover}
\label{s.app1}

\begin{figure}[htb!]
 \centering
\includegraphics[width=0.9\columnwidth]{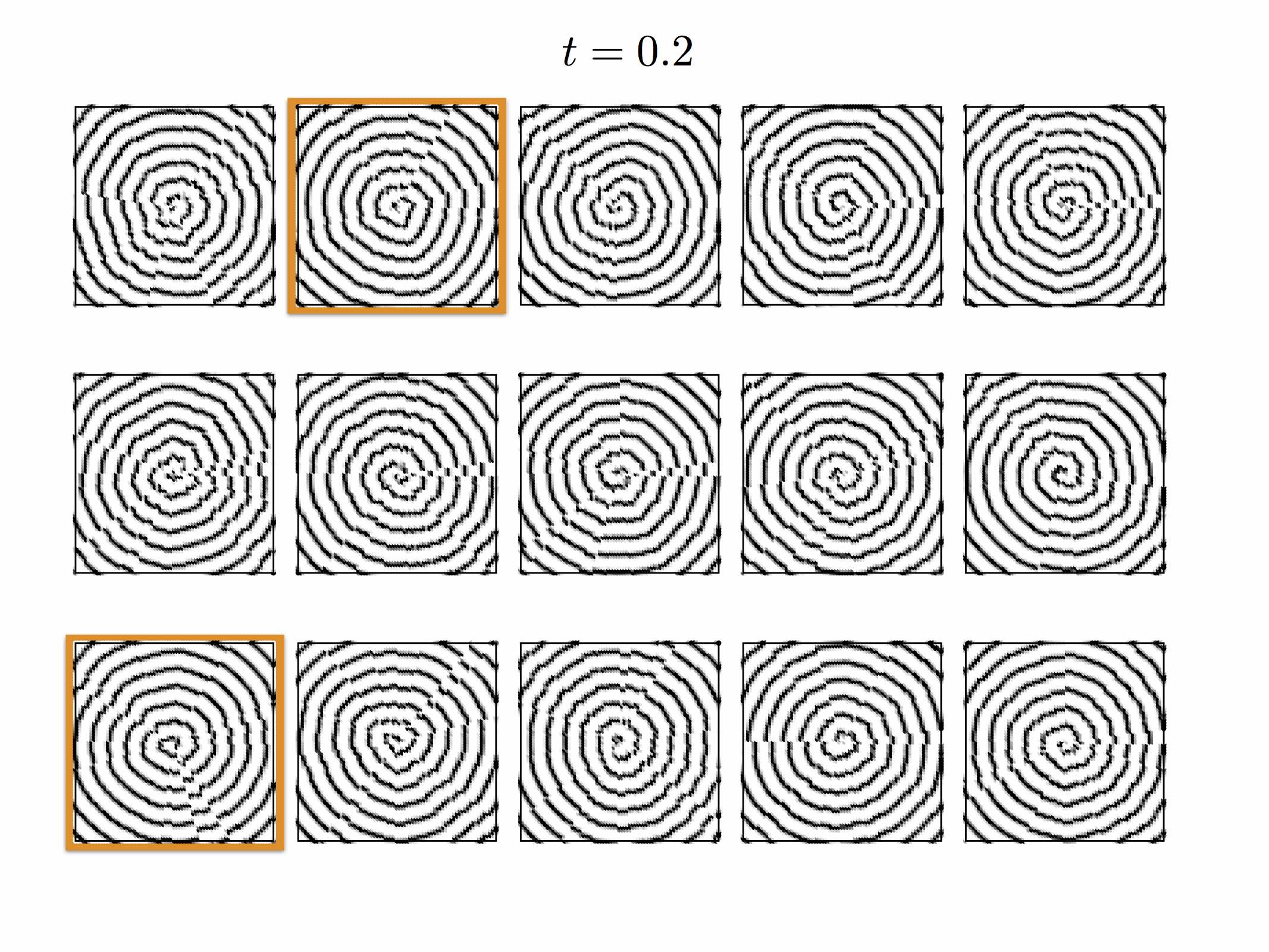}
 \caption{Representative interference patterns reconstructed from snapshots of the MC simulations on a $L=48$ ring with $u=0.25$ and $t=0.2$. The framed patterns correspond to path-integral configurations displaying quantum fluctuations of the winding number (namely a $\tau$-dependent $W$).}
\label{f.t0.2}
\end{figure}

\begin{figure}[htb!]
 \centering
\includegraphics[width=0.9\columnwidth]{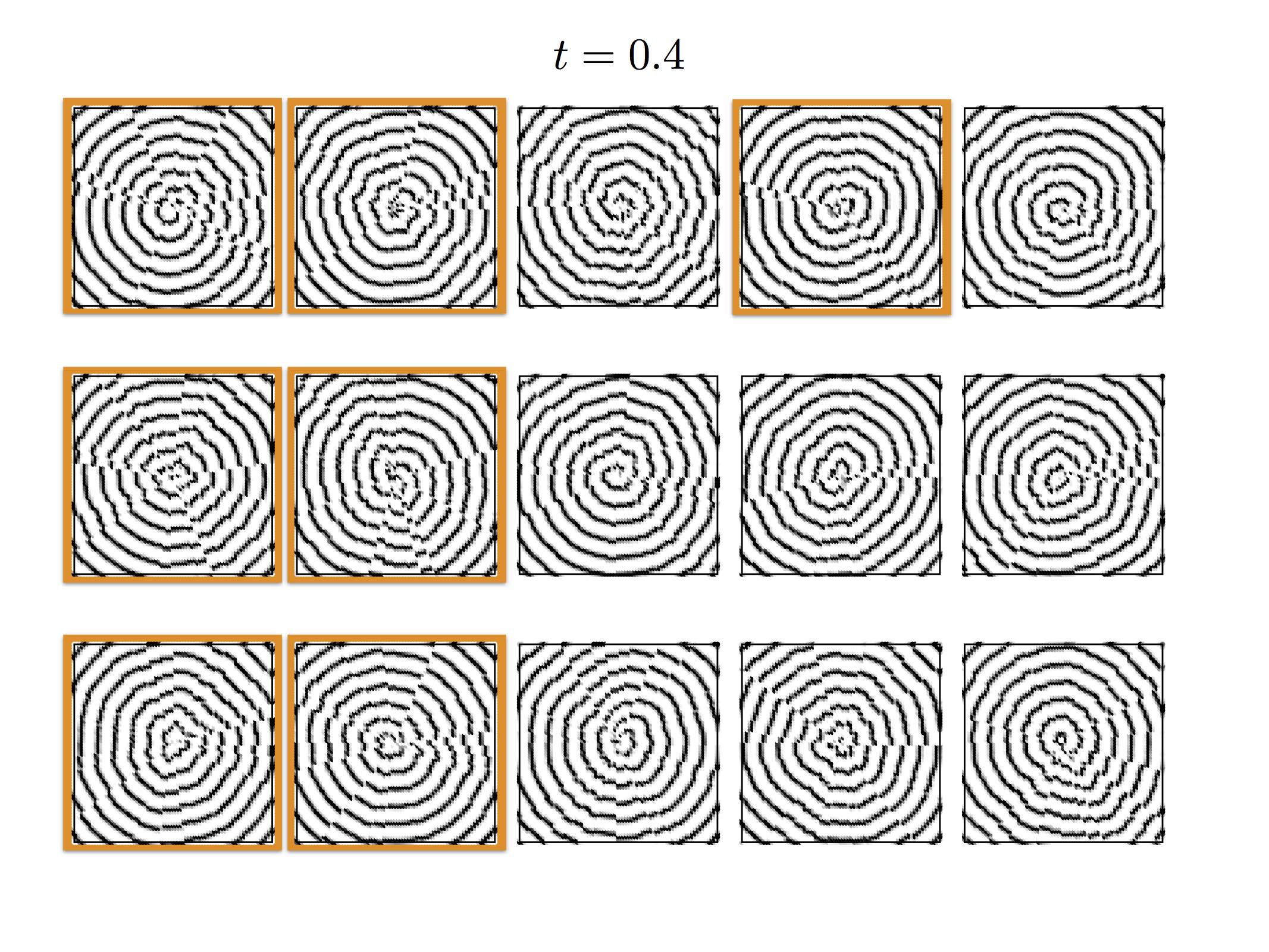}
 \caption{Same as in Fig.~\ref{f.t0.2}, but with $t=0.4$.}
\label{f.t0.4}
\end{figure}

\begin{figure}[htb!]
 \centering
\includegraphics[width=0.9\columnwidth]{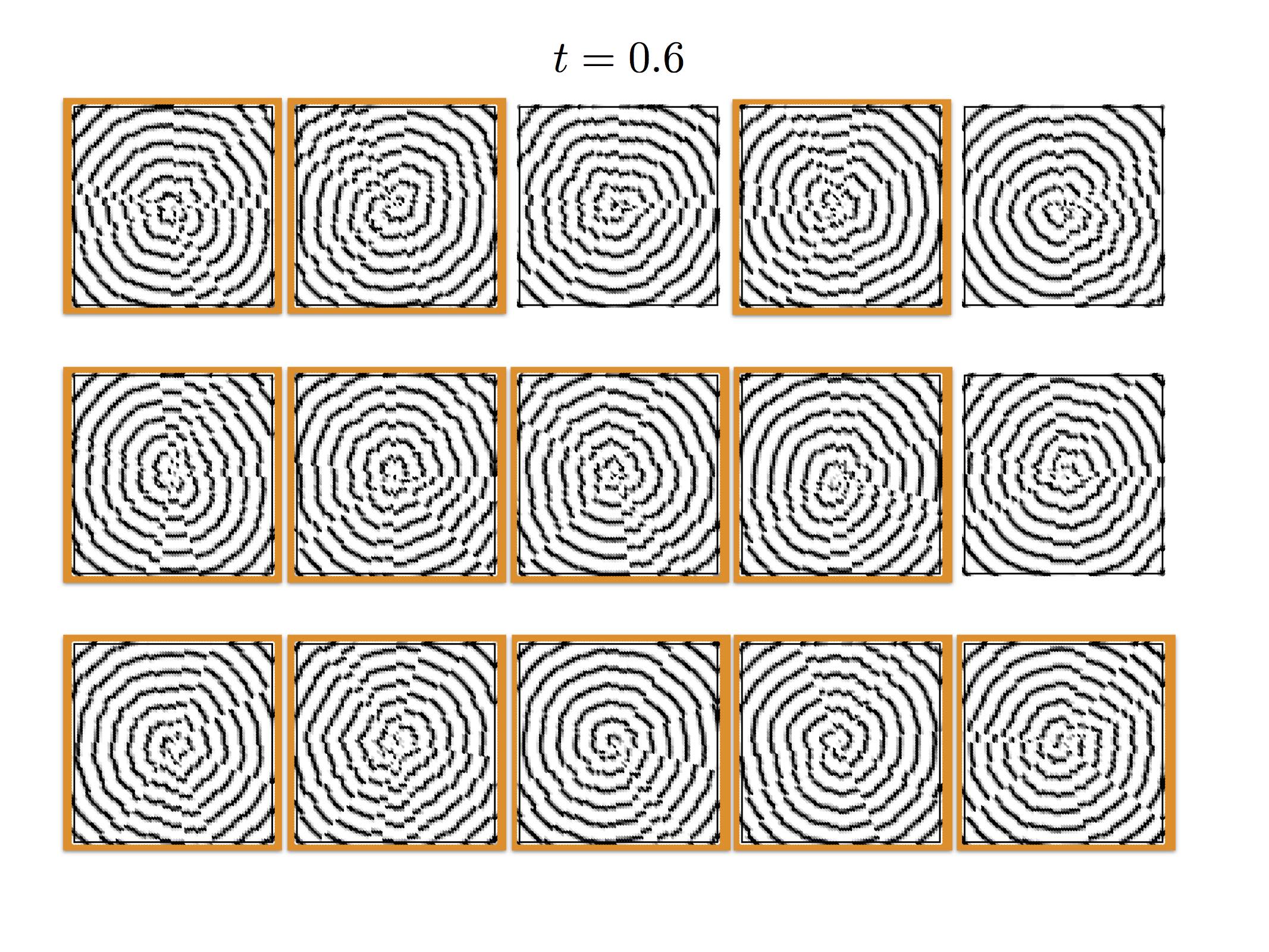}
 \caption{Same as in Fig.~\ref{f.t0.2}, but with $t=0.6$.}
\label{f.t0.6}
\end{figure}

In this appendix we present representative interference patterns with $|W|=1$, stemming from uncorrelated snapshots of our Monte Carlo simulations on a $L=48$ ring with weak interactions $u=0.25$. Such patterns mimick the outcome of independent experimental runs obtained under the same conditions. Figs.~\ref{f.t0.2}, ~\ref{f.t0.4} and ~\ref{f.t0.6} show interference patterns for a growing temperature, and they establish a link between the appearance of localized dislocations in the interference fringes, and their origin from quantum-coherent fluctuations of the winding number.  

\bibliographystyle{jphysicsB}
\bibliography{Bosering}

\end{document}